\newif{\ifjournal}
  \journaltrue

\ifjournal
  \documentclass{mn2e}
  \usepackage{graphicx,mathptm}
  \newcommand{\gtrsim}{\ga}
  \newcommand{\lesssim}{\la}
\else
  \documentclass{paper}
  \newcommand{\ga}{\gtrsim}
  \newcommand{\la}{\lesssim}
\fi

\newcommand{\be}{\begin{equation}}
\newcommand{\ee}{\end{equation}}
\newcommand{\bed}{\begin{displaymath}}
\newcommand{\eed}{\end{displaymath}}
\newcommand{\ba}{\begin{eqnarray}}
\newcommand{\ea}{\end{eqnarray}}
\newcommand{\nn}{\nonumber \\}

\begin{document}
\input{psfig.sty}
\title[A dynamical model for the distribution of dark matter and gas in
galaxy clusters] 
{A dynamical model for the distribution of dark matter and gas in
galaxy clusters}

\author[Rasia et al.]
{Elena Rasia$^1$\thanks{rasia@pd.astro.it}, Giuseppe Tormen$^1$,
Lauro Moscardini$^2$\\
$^1$Dipartimento di Astronomia, Universit\`a di Padova,
vicolo dell'Osservatorio 2, I-35122 Padova,
Italy \\
$^2$Dipartimento di Astronomia, Universit\`a di Bologna, 
via Ranzani 1, I-40127 Bologna, Italy}

\maketitle

\begin{abstract}

Using the results of an extended set of high-resolution non-radiative
hydrodynamic simulations of galaxy clusters we obtain simple analytic
formulae for the dark matter and hot gas distribution, in the
spherical approximation. Starting from the dark matter phase-space
radial density distribution, we derive fits for the dark matter
density, velocity dispersion and velocity anisotropy.  We use these
models to test the dynamical equilibrium hypothesis through the Jeans
equation: we find that this is satisfied to good accuracy by our
simulated clusters inside their virial radii. This result also show
that our fits constitute a self-consistent dynamical model for these
systems.

We then extend our analysis to the hot gas component, obtaining
analytic fits for the gas density, temperature and velocity structure,
with no further hypothesis on the gas dynamical status or state
equation.  Gas and dark matter show similar density profiles down
to $\approx 0.06 R_{\rm v}$ (with $R_{\rm v}$ the virial radius),
while at smaller radii the gas flattens producing a central core.  Gas
temperatures are almost isothermal out to roughly $0.2 R_{\rm v}$,
then steeply decrease reaching at the virial radius a value almost a
factor of two lower.  We find that the gas is not at rest inside
$R_{\rm v}$: velocity dispersions are increasing functions of the
radius, motions are isotropic to slightly tangential, and contribute
non-negligibly to the total pressure support.  We test this model
using a generalization of the hydrostatic equilibrium equation, where
the gas motion is properly taken into account.  Again we find that the
fits provide an accurate description of the system: the hot gas is in
equilibrium and is a good tracer the overall cluster potential if all
terms (density, temperature and velocity) are taken into account,
while simpler assumptions cause systematic mass underestimates. In
particular, we find that using the so-called $\beta$-model
underestimates the true cluster mass by up to 50 per cent at large
radii.  We also find that, if gas velocities are neglected, then a
simple isothermal model fares better at large radii than a
non-isothermal one.

The shape of the gas density profile at small radii is at least
partially explained by the gas expansion caused by energy transfer
from dark matter during the collapse.  In fact, when gas bulk energy
is also considered, gas and dark matter are in energy equipartition in
the final system at radii $r > 0.1 R_{\rm v}$, while at smaller radii
the gas is hotter than the dark matter.  This energy unbalance is also
probably the reason of the further global halo compression compared to
a pure collisionless collapse, which we point out by comparing the
dark matter and total density profiles of our hydro-simulated clusters
with a set of identical - but pure N-body - ones.  The compression has
the effect of rising the mean concentration by an amount of roughly 10
per cent.
\end{abstract}


\section{Introduction}

Galaxy clusters are the largest virialized systems we observe in the
universe today.  To first order they can be described as dark matter
(DM) haloes, since dark matter makes up at least 80 per cent of their
total mass.  The second most important component of galaxy clusters is
a hot plasma, called the intracluster medium (ICM); it has temperature
of the order of a few keV, roughly tracing the depth of the
gravitational potential wells of the systems.

Any comparison with observations in the optical or via gravitational
lensing requires modelling at least the dark matter component of
galaxy clusters, while X-ray or millimetric observations [e.g. the
Sunyaev-Zel'dovich (SZ) effect] require knowledge of the internal
structure of the ICM.  Both aspects can in principle be studied by
mean of numerical simulations.  In fact, in the last few years
simulations have reached enough resolution to reliably describe the
structural properties of DM haloes down to the kpc scale.  An
important result along this line is the discovery that relaxed dark
matter haloes exhibit a density profile which can be accurately
described by a double power-law, with inner asymptotic slope of around
$-1$ and outer asymptotic slope close to $-3$ (Navarro, Frenk \& White
1996, 1997, hereafter NFW; Cole \& Lacey 1996; Tormen, Bouchet \&
White 1997).  Although the exact value of the exponent of the inner
slope is still somewhat debated (Moore et al. 1998; Jing \& Suto 2000;
Jing 2000; Ghigna et al. 2000; Power et al. 2003), the overall model
is quite well defined and robust.  Fitting formulae for the density
profile of galaxy clusters are very useful to probe the standard
paradigm of structure formation and cosmological models against
observations, like the distribution of galaxies in clusters (e.g. the
CNOC project: Carlberg et al. 1997), and the gravitational lensing
properties of galaxy clusters (e.g. Bartelmann et al. 1998;
Meneghetti, Bartelmann \& Moscardini 2003).

However, density profiles contain information only on the structural
properties of dark matter haloes, while knowledge of the velocity
structure is required if one wants to extend the description to the
dynamical properties.  A few attempts have been done in this
direction, both using analytical calculations and numerical
simulations (e.g. Sheth et al. 2001; Taylor \& Navarro 2001; Hiotelis
2002a,b).  These papers usually made simplifying assumptions on the
dynamics (e.g. isotropic velocity dispersion or a given functional
form for the phase-space density or for the velocity anisotropy) and
imposed the Jeans equation as a constraint to derive the model
equations.  Still, the difficulty of the problem has made it
impossible to generate so far a self-consistent analytical model of
the density and velocity structure of dark matter haloes with no
assumption on the density or velocity field besides spherical
symmetry.  An easier - but equally interesting - way is that of taking
the answer directly from numerical simulations, and use the mean
radial profiles of a sample of simulated clusters as a first guess for
such dynamical model, with no assumption on the dynamical status of
the system.  This is the first goal of the present paper.

The second, and perhaps even more interesting goal, is extending this
attempt to modelling the hot intracluster gas.  Given a double
power-law functional form for the dark matter density profile, it is
possible to calculate the radial distribution of the hot gas by
imposing the hydrostatic equilibrium equation and assuming that the
gas is either isothermal or polytropic (e.g. Suto, Sasaki \& Makino
1998; Komatsu \& Seljak 2001; Ascasibar et al. 2003, hereafter
AYMG03).  Very recently, Lee \& Suto (2003) presented a physical model
for the non-spherical distribution of the intracluster gas in
hydrostatic equilibrium under the gravity of triaxial dark matter
halos (Jing \& Suto 2002): their solutions for the gas and temperature
profiles have been obtained in both isothermal and polytropic
equations of state.  In the present work we instead take the approach
of deriving all relevant quantities directly from the simulations,
with no assumptions on the gas state equation or dynamical
equilibrium.  We measure the mean radial gas density and temperature
profiles of our simulated clusters and model them using analytical
fits.  As the gas in simulated clusters is not completely at rest
inside the virial radius, we also derive fits for the gas velocity
dispersions and anisotropy.  The fits are then tested against the
hydrostatic equilibrium equation - and not imposing it.  The resulting
model gives us an idea of how relaxed are, on average, simulated
clusters and how well they are described by the equilibrium equations.
We believe that it is important to have a complete model for ICM+DM
because observations measure the total mass distribution, and the
presence of gas, even if not dominant, can well affect the DM
component.

The plan of the paper is as follows. In Section 2 we present the
hydrodynamical N-body simulations used in this work.  In Section 3 we
model the dark matter density and the velocity structure, and then use
our model to test the dynamical equilibrium of clusters through the
Jeans equation, cast in the form of a mass estimator.  In Section 4 we
perform a similar analysis for the hot gas component.  In Section 5 we
estimate the effect of the presence of ICM on the density profile.
Moreover we discuss the dependence of our results on the dynamical and
environmental status of galaxy clusters and we compare our results to
previous work.  Finally, our conclusions are summarized in Section 6.

\section {Numerical Simulations}

In order to produce our sample of simulated galaxy clusters, we used
the re-simulation technique, based on a software package developed by
our group (ZIC: Zoom Initial Conditions).  The method is presented in
Tormen et al.  (1997).  Here we give only a brief summary of the
adopted procedure.

We started from a cosmological N-body simulation (the `parent
simulation'), with $512^3$ particles in a box of side $479 h^{-1}$
Mpc.  This has been produced by N. Yoshida for the Virgo Consortium
(Yoshida, Sheth \& Diaferio 2001; see also Jenkins et al. 2001).  The
assumed cosmological model is a flat universe, where the contributions
to the density parameter from dark matter, baryons and cosmological
constant are $\Omega_{\rm DM} = 0.27$, $\Omega_{\rm b}=0.03$ and
$\Omega_\Lambda = 0.7$, respectively.  The value of the Hubble
constant (in units of 100 km/s/Mpc) is $h = 0.7$.  The initial
conditions correspond to a cold dark matter power spectrum such that
the r.m.s. matter fluctuation in sphere with radius of $8 h^{-1}$ Mpc
is $\sigma_8 = 0.9$.  Each particle has a mass $6.86 \times 10^{10}
h^{-1} M_\odot$, which allows to resolve cluster-sized haloes by
several thousand particles. The gravitational softening was fixed at
$30 h^{-1}$ kpc.

\begin{table*}
\caption{Main characteristics of the simulated clusters. Column 1:
identification number. Columns 2 and 3: number of dark matter ($N_{\rm
DM}$) and gas particles ($N_{\rm gas}$) inside the virial radius
$R_{\rm v}$.  Columns 4 and 5: mass of the dark matter ($M_{\rm DM}$)
and gas ($M_{\rm gas}$) components inside the virial radius.  Columns
6 and 7: total (i.e. virial) mass $M_{\rm v}$ and virial radius
$R_{\rm v}$.}
\begin{center}
\begin{tabular}{|c||c|c|c|c|c|c|}
~~\\ Number & $N_{\rm DM}$ & $N_{\rm gas}$ & $M_{\rm DM}(h^{-1}
M_{\odot})$ & $M_{\rm gas} (h^{-1} M_{\odot})$ & $M_{\rm v}
(h^{-1} M_{\odot})$ & $R_{\rm v}(h^{-1}$ Mpc)\\
\hline
1 &282574   &262319    &7.633$\times 10^{14}$   &7.873$\times 10^{13}$
&8.420$\times 10^{14}$ &1.953 \\
2 &278569   &257953    &1.254$\times 10^{15}$   &1.291$\times 10^{14}$
&1.383$\times 10^{15}$ &2.305 \\
3 &85159    &79644     &3.834$\times 10^{14}$   &3.985$\times 10^{13}$
&4.233$\times 10^{14}$ &1.553 \\
4 &294373   &272001    &1.325$\times 10^{15}$   &1.361$\times 10^{14}$
&1.461$\times 10^{15}$ &2.347 \\
5 &179681   &168983    &8.082$\times 10^{14}$   &8.446$\times 10^{13}$
&8.927$\times 10^{14}$ &1.991 \\
6 &146386   &137365    &6.585$\times 10^{14}$   &6.865$\times 10^{13}$
&7.271$\times 10^{14}$  &1.860 \\
7 &318653   &292996    &1.087$\times 10^{15}$   &1.111$\times 10^{14}$
&1.198$\times 10^{15}$ &2.197 \\
8 &427583   &413492    &1.538$\times 10^{15}$   &1.653$\times 10^{14}$
&1.703$\times 10^{15}$ &2.470 \\
9 &166855   &135641    &6.002$\times 10^{14}$   &5.421$\times 10^{13}$
&6.544$\times 10^{14}$  &1.796 \\
10 &275259   &244874    &4.960$\times 10^{14}$   &4.903$\times 10^{13}$
 &5.450$\times 10^{14}$ &1.691 \\
11 &158345   &146466    &2.853$\times 10^{14}$   &2.932$\times 10^{13}$
&3.146$\times 10^{14}$ &1.407 \\
12 &190453   &174358    &6.860$\times 10^{14}$   &6.978$\times 10^{13}$
 &7.558$\times 10^{14}$ &1.884 \\
13 &101482   &95697     &3.656$\times 10^{14}$   &3.830$\times 10^{13}$
 &4.039$\times 10^{14}$  &1.529 \\
14 &159330   &149170    &7.169$\times 10^{14}$   &7.458$\times 10^{13}$
 &7.915$\times 10^{14}$ &1.913 \\
15 &107229   &96981     &4.825$\times 10^{14}$   &4.849$\times 10^{13}$
 &5.310$\times 10^{14}$ &1.675 \\
16  &58734  &57358     &3.606$\times 10^{14}$   &3.913$\times 10^{13}$
  &3.997$\times 10^{14}$  &1.524 \\
17  &71937  &66900     &4.417$\times 10^{14}$   &4.564$\times 10^{13}$
 &4.873$\times 10^{14}$  &1.628 \\
\end{tabular}
\label{tab:char}
\end{center}
\end{table*}

We identified all collapsed objects from the $z=0$ output of the
parent simulation, using a spherical overdensity criterion.  From this
catalogue we randomly selected a number of cluster-sized dark matter
haloes and carved around each of them a spherical region encompassing
a few virial radii.  Each of these regions was then resampled to build
new initial conditions suitable for a higher number of particles -
using on average $10^6$ dark matter particles and the same number of
gas particles.  Outside this `high resolution' region the number of
particles was strongly reduced by interpolating the original particles
onto a coarse spherical grid with $3$ degree angular resolution. This
produced a number of `low resolution' macro particles which still give
a good representation of the large-scale tidal field necessary to form
the cluster in the high resolution region.  Particle masses in the
high-resolution region ranged from $2\times 10^9 h^{-1} M_\odot$ to
$6\times 10^9 h^{-1} M_\odot$ for dark matter, and from $3\times 10^8
h^{-1} M_\odot$ to $7\times 10^8 h^{-1} M_\odot$ for gas.

We finally evolved these new initial conditions using the publicly
available code GADGET (GAlaxies with Dark matter and Gas intEracT;
Springel et al. 2001b).  This is a TREESPH code where the dark matter
particles are evolved using a Barnes \& Hut (1986) tree-code, while
the collisional gas is followed using a Smoothed Particle
Hydrodynamics (SPH) approach. The re-simulations, which include only
non-radiative hydrodynamics, started at redshift $z_{\rm in} = 35-50$,
depending on the mass resolution.  Gravitational softening was fixed
at $5 h^{-1}$ kpc (cubic spline).

The final sample includes 17 high-resolution clusters, with virial
masses ranging between $3.1 \times 10^{14} h^{-1} M_{\odot}$ and
$1.7\times 10^{15} h^{-1} M_{\odot}$ and virial radii between 1.4 and
$2.5 h^{-1}$ Mpc.  Virial radii were defined using the overdensity
threshold dictated by the spherical top-hat model (e.g. Eke et
al. 1996).

In Table~\ref{tab:char} the main properties of the cluster sample are
summarized.  Our analysis is restricted to the $z=0$ outputs.  A
discussion of the dynamical properties of cluster substructure done
using the results of these simulations is presented elsewhere (Tormen,
Moscardini \& Yoshida 2004). As a final comment, we stress that the
random criterion used to select the re-simulated haloes leads to a
sample of clusters with varying dynamical properties: at the present
time, some are more relaxed, while others are dynamically perturbed.
The surrounding environment can also be quite different: some clusters
are more isolated, while others are interacting with the surrounding
cosmic web.  This means that the modelling we will do will be
representative of an average cluster, in an average environment and
dynamical configuration.

\section{Modelling Dark Matter Profiles}

As already discussed in the introduction, the underlying idea of this
work is to build a self-consistent model for the radial profiles
describing the dynamical properties of galaxy clusters.  The general
criterion we followed is to use at most double power-law analytic fits
to all differential profiles.  This was done in order to keep the
number of free parameters to the minimum, and to allow easily
integration to obtain e.g. the cluster's mass.

\subsection{Coarse-grained distribution function}

Taylor \& Navarro (2001) have recently noticed that in N-body
simulations of galaxy-sized dark matter haloes the radial profile of
the coarse-grained distribution function, $\overline{f}(x) \equiv
\rho(x)/\sigma_r^3(x)$,  is well fitted by a single power-law:
\be
\label{eq:phspace}
\overline{f}(x) \propto x^{\alpha}\ ,
\ee
with $\alpha\approx -1.875$.  We recall that $\rho(x)$ and
$\sigma_r(x)$ are the differential dark matter density and the radial
velocity dispersion, respectively, at distance $x \equiv r/R_{\rm v}$
from the cluster centre, where $R_{\rm v}$ is the cluster virial
radius.  The quantity $\overline{f}(x)$ has the dimension of a
phase-space density, and can be thought of as the coarse-grained
phase-space density profile of a halo.  A power-law behaviour for
$\overline{f}(x)$ is also expected in the self-similar solution for
secondary infall of gas onto a point-mass perturber in a uniformly
expanding universe (Bertschinger 1985).  Since a single power-law is
an easy profile to fit, we deemed it a good starting point for our
investigation.

 \begin{figure} \centering
\includegraphics[width=7.cm]{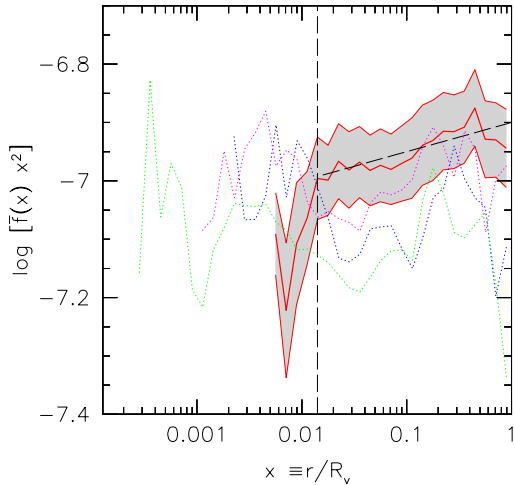}
\caption{The coarse-grained phase-space density mean radial profile for
dark matter in our simulations.  To enhance clarity, we plot
$\overline{f} x^2$ (solid line) and its error (shaded region). The
dashed line is our best single power-law fit
(equation~\ref{eq:phspace}).  The three dashed noisy lines represent
the profiles of a single cluster re-simulated - with dark matter only
- at different increasing resolutions (notice the different extension
of the lines to the left).  The vertical dashed line was taken as the
limit of our model, and corresponds to $\log x= -1.85$.  At this
distance the mean profile starts to show fluctuations originated by
numerical artifacts.}
\label{fig:phaserho}
\end{figure}

For each cluster we computed the radial profile of $\overline{f}(x)$
versus the distance $x$ from the cluster centre.  We then averaged the
profiles and fit this to a power-law relation.  The result is shown in
Fig.~\ref{fig:phaserho}: the solid line and the surrounding shaded
region represent the mean profile with its error.

From the figure we can notice that a single power-law fit as in
equation~(\ref{eq:phspace}) is indeed a good description of the data
at most radii.  Our best-fitting slope, $\alpha=-1.95$, is slightly
steeper than that obtained by Taylor \& Navarro (2001).  However, at
scales corresponding to $\log x\la -1.85$, i.e. for $x\la0.014$, the
mean profile starts showing significant fluctuations from the fit.
While a departure from the single power-law is expected for $x \to 0$
(as at small radii the density diverges, but the radial velocity
dispersion goes to zero), the observed deviation is probably due to
numerical effects.

In order to confirm or dispute this interpretation, we computed the
analogous profile for a single cluster, re-simulated with dark matter
only, at higher and higher resolution (see Springel et al. 2001a).
The resulting profiles are shown in Fig.~\ref{fig:phaserho} by the
three dotted lines: they correspond to simulations with $6\times
10^5$, $3.4\times 10^6$ and $1.9\times 10^7$ particles inside the
virial radius at redshift $z=0$.  These profiles, which are plotted
from the radius containing 250 particles, are quite noisy because they
refer to a single cluster realisation as opposed to our mean curve.
However, their slope is quite close to that obtained from our sample,
essentially confirming our estimation of $\alpha$.  More importantly,
the three curves show no systematic deviation from a single power-law
at small radii: this seems to confirm the numerical origin of the
small scale fluctuations.  We therefore decided to consider all the
profiles presented in this work only at radii larger than $\log x\la
-1.85$.  Within this radius our clusters contain on average $\approx
800$ dark matter particles.  This is a quite conservative limit, and
satisfies the requirements suggested by recent convergence studies
(Power et al. 2003).

Notice that the normalisation of the power-law relation for our fit is
slightly larger than that resulting from the single high-resolution
dark matter-only simulations. This difference could either be due to
statistical fluctuations or by the fact that our simulations contain
also gas.  However, the discrepancy is not really relevant for our
model, as in what follows we will only use the slope $\alpha$ of the
relation.

\subsection{Velocity structure}

As a next step we will consider the dark matter velocity dispersion
profiles.  These are important not only to understand the velocity
structure of dark matter haloes, but also to estimate the total
cluster mass through the Jeans equation (as done in section
\ref{sec:dm_mass_est}).

\begin{figure}
\centering
\includegraphics[width=7.cm]{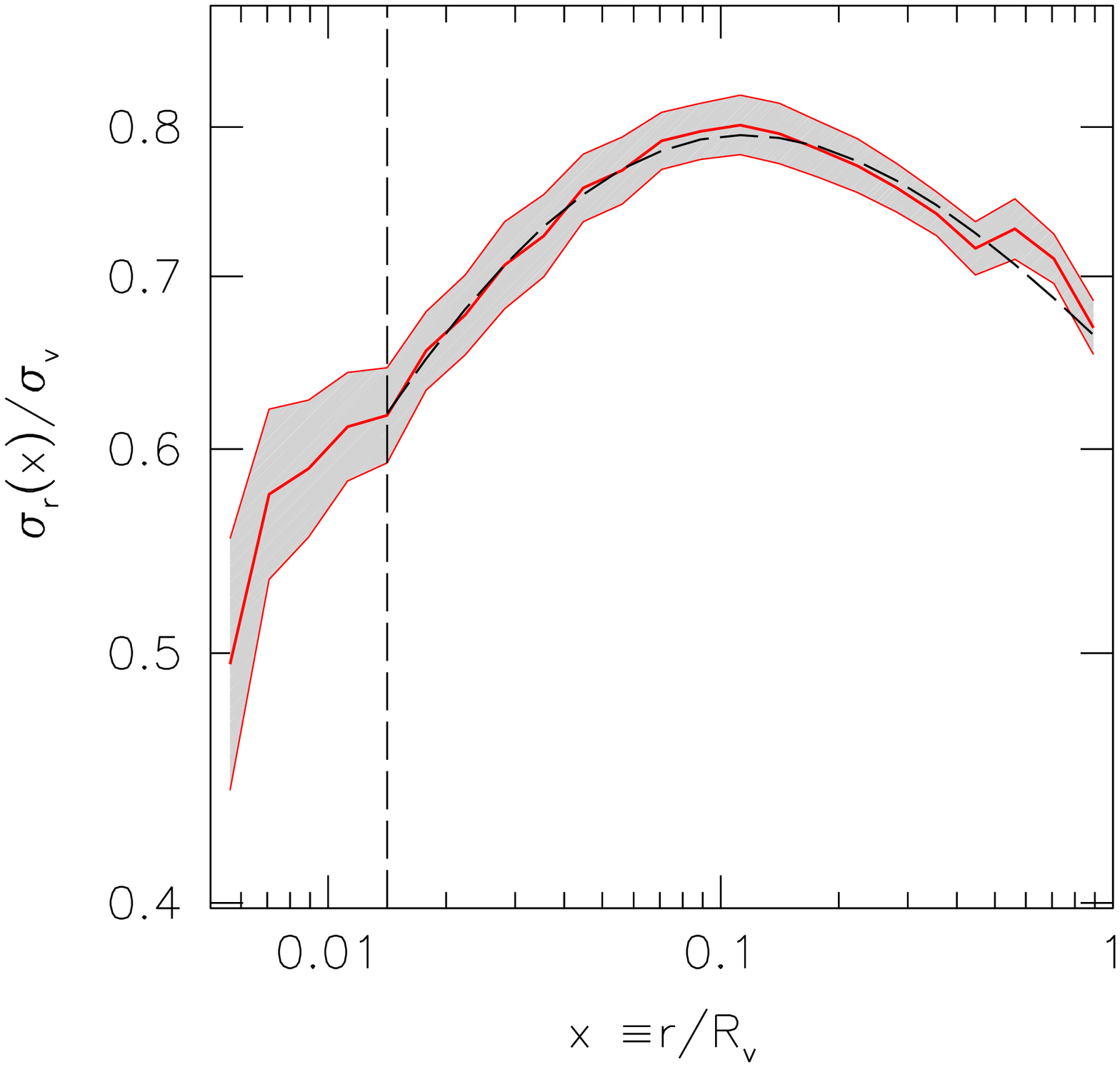}
\includegraphics[width=7.cm]{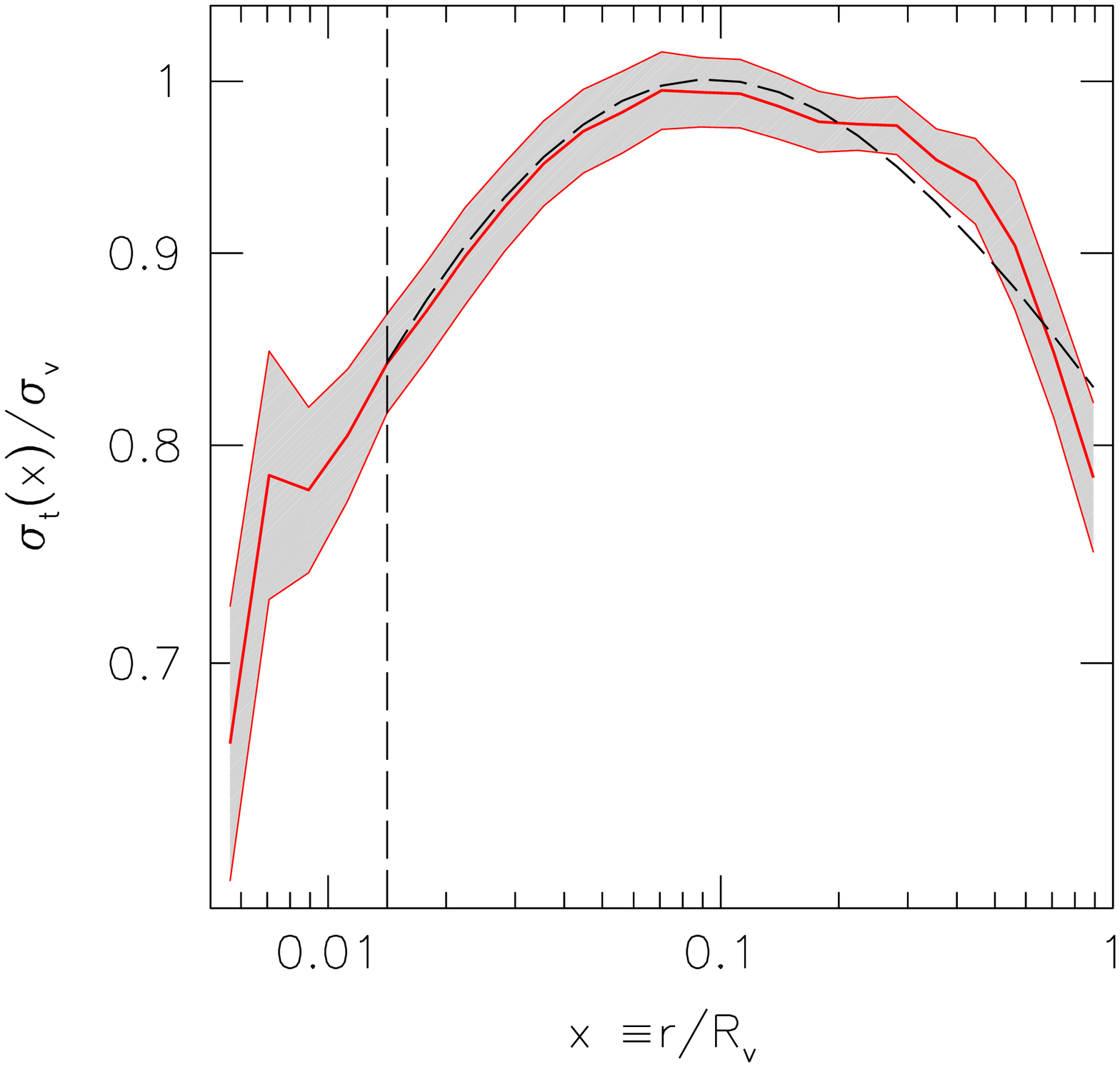}
\includegraphics[width=7.cm]{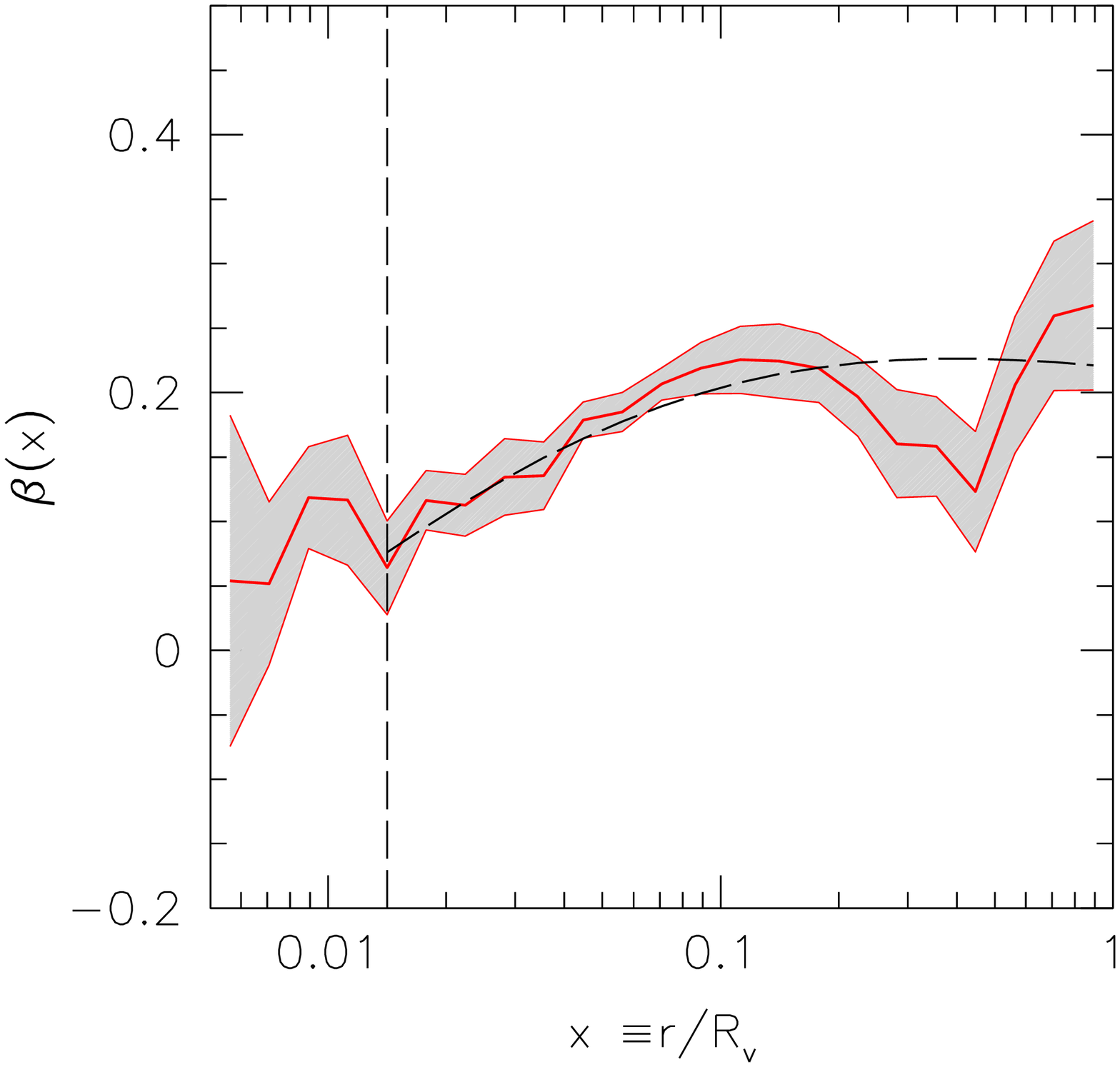}
\caption{
Velocity profiles. From top to bottom, different panels refer to
radial velocity dispersion $\sigma_r(x)$, tangential velocity
dispersion $\sigma_t(x)$, and velocity anisotropy $\beta(x)$.
Velocity dispersions are plotted in units of the virial quantity
$\sigma_{\rm v}$.  The solid lines and surrounding shaded regions
represent the mean profile obtained from our sample of simulated
clusters and its error, respectively; dashed lines represent our
fitting formulae (equations~\ref{eq:sr}-\ref{eq:st}; see text for more
details).  The vertical dashed line is the limit of our model, set at
$\log x= -1.85.$}
\label{fig:sigdm}
\end{figure}

In Fig.~\ref{fig:sigdm} we show the mean profiles (averaged over our
cluster sample) for the radial velocity dispersion $\sigma_r(x)$ (top
panel), the tangential velocity dispersion $\sigma_t(x)$ (central
panel) and the velocity anisotropy parameter $\beta(x) \equiv
1-\sigma_t^2(x)/2\sigma_r^2(x)$ (bottom panel).  The solid bands
represent the error on the mean.  In order to perform the mean over
clusters of different mass, each profile was normalized by the
quantity $\sigma_{\rm v}^2 \equiv GM_{\rm v}/R_{\rm v}$, where $R_{\rm
v}$ and $M_{\rm v}$ are the individual virial radii and masses. The
vertical dashed line represents the limit of our model, at $\log x=
-1.85$.  We remind that the velocity anisotropy $\beta$ shows
preferred directions in the velocity field.  In particular, $\beta=0$
corresponds to an isotropic velocity field, while $\beta<0$ and $0<
\beta \leq 1$ indicate predominance of tangential and radial motions, 
respectively.

The plots show that in general the dark matter velocity field in
clusters is not isothermal nor isotropic. The radial and tangential
velocity dispersions decline both at small and large radii, while the
radial trend of $\beta$ indicates a predominance of radial motions at
large radii, and an almost isotropic velocity field in the inner part,
a result consistent with previous studies (e.g. Cole \& Lacey 1996;
Tormen et al. 1997).  We also find that the mean radial velocity (not
shown) is still negative at the virial radius, i.e. infall motions are
present: this suggests that our clusters are on average not completely
relaxed, as clumps of matter are still entering at $z=0$.

Since the three quantities $\sigma_r$, $\sigma_t$ and $\beta$ are not
independent, we can derive any of them using the other two.  This
should also be true for the analytic fits; therefore we tried fitting
each pair of profiles with double power-laws, and derived each third
fit by their combination.  The solution which best describes the
simulated profiles turned out to be fitting the radial velocity
dispersion and the velocity anisotropy directly, and getting the
tangential velocity dispersion by combination.

The analytical expressions of our fits (hereafter marked by a tilde)
are the following:

\begin{enumerate}

\item
radial velocity dispersion:
\be
\label{eq:sr}
\tilde{\sigma}_{r}(x) \equiv \sigma_r(x) / \sigma_{\rm v}
 = \sigma_{r_0} x^{0.3}(x+x_{p_1})^{-0.48},    
\ee
with scale radius $x_{p_1}=10^{-1.16}$ and normalization amplitude
$\sigma_{r_0}=0.67$;
 
\item
velocity anisotropy $\beta(x)$:
\be
\label{eq:b}
\tilde{\beta}(x)= \beta_0 x^{0.25}(x+x_{p_2})^{-0.3},
\ee
with $x_{p_2}=10^{-1.1}$ and $\beta_0=1.7$;

\item  
tangential velocity dispersion:
\ba
\label{eq:st}
\tilde{\sigma}_t(x) & \equiv & \sigma_t(x) / \sigma_{\rm v} \nn
&=&\tilde{\sigma_r}(x)\sqrt{2 \left[ 1-\tilde{\beta}(x)\right]}
= \sigma_{r_0} x^{0.3} \left( x + x_{p_1}\right)^{-0.48} \nn
&\times& \sqrt{ 2\left[ 1 - \beta_0 x^{0.25}(x + x_{p_2})^{-0.3}\right]}\ .
\ea

\end{enumerate}

\begin{figure}
\centering
\includegraphics[width=7.cm]{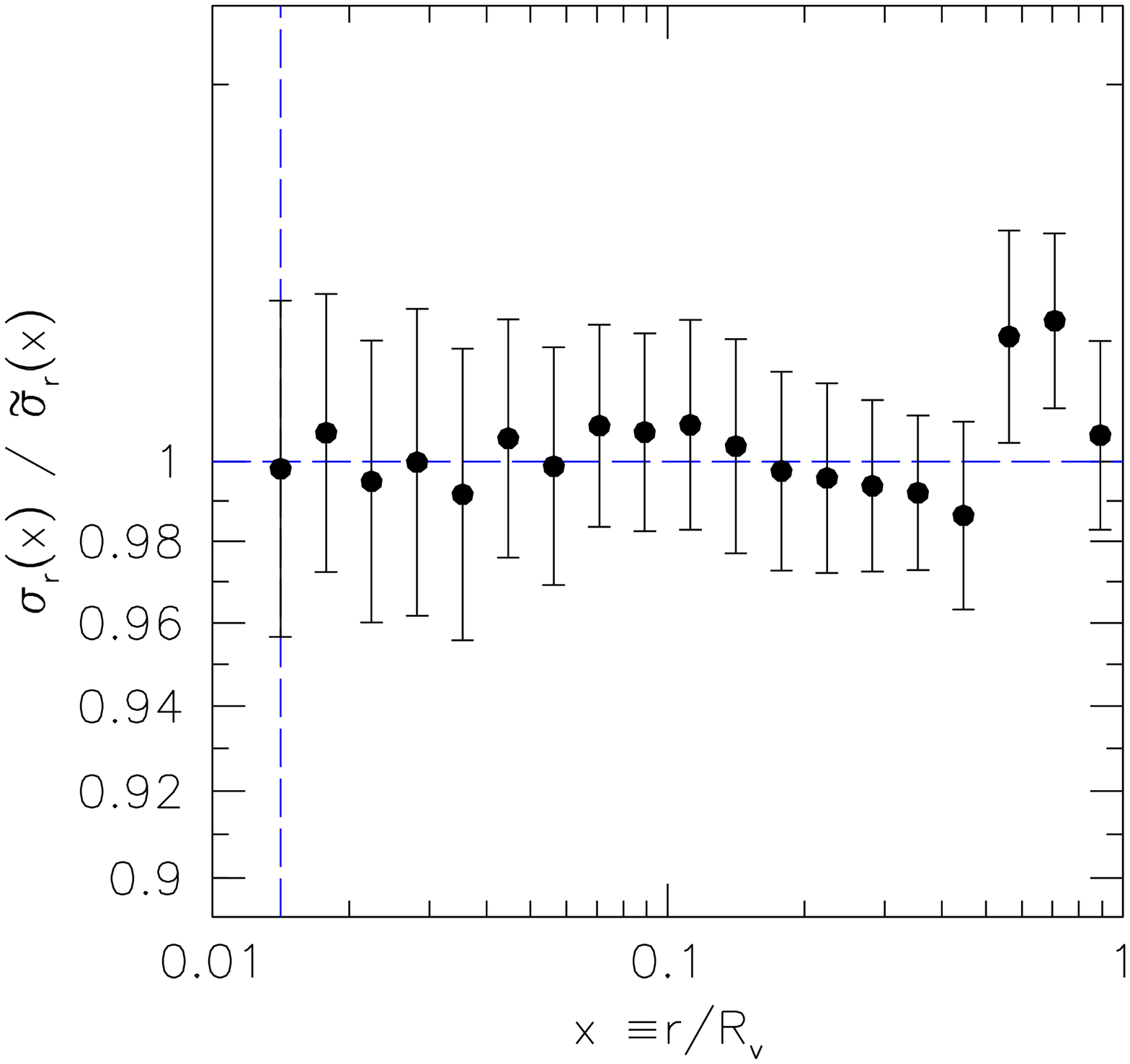}
\includegraphics[width=7.cm]{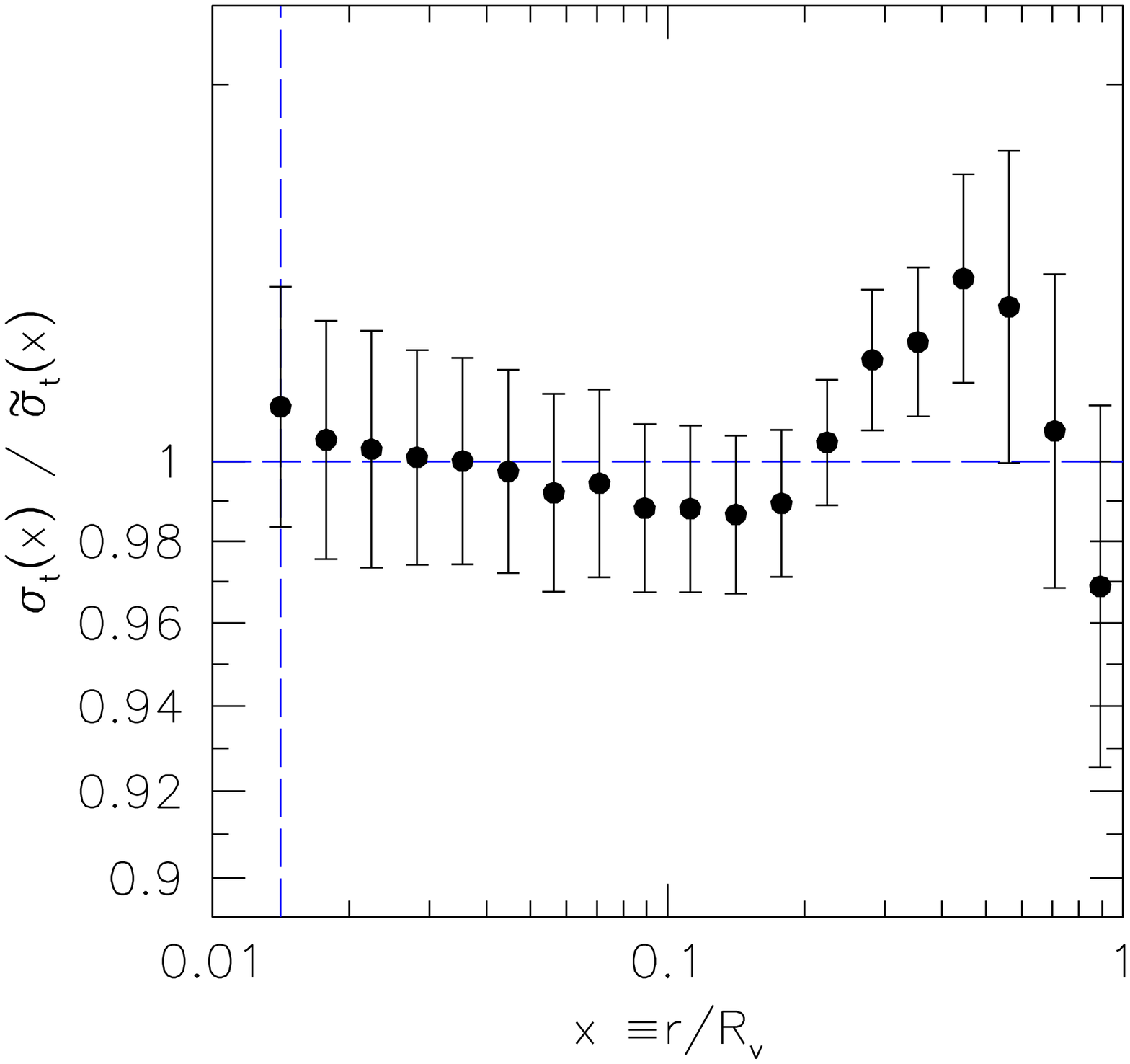}
\includegraphics[width=7.cm]{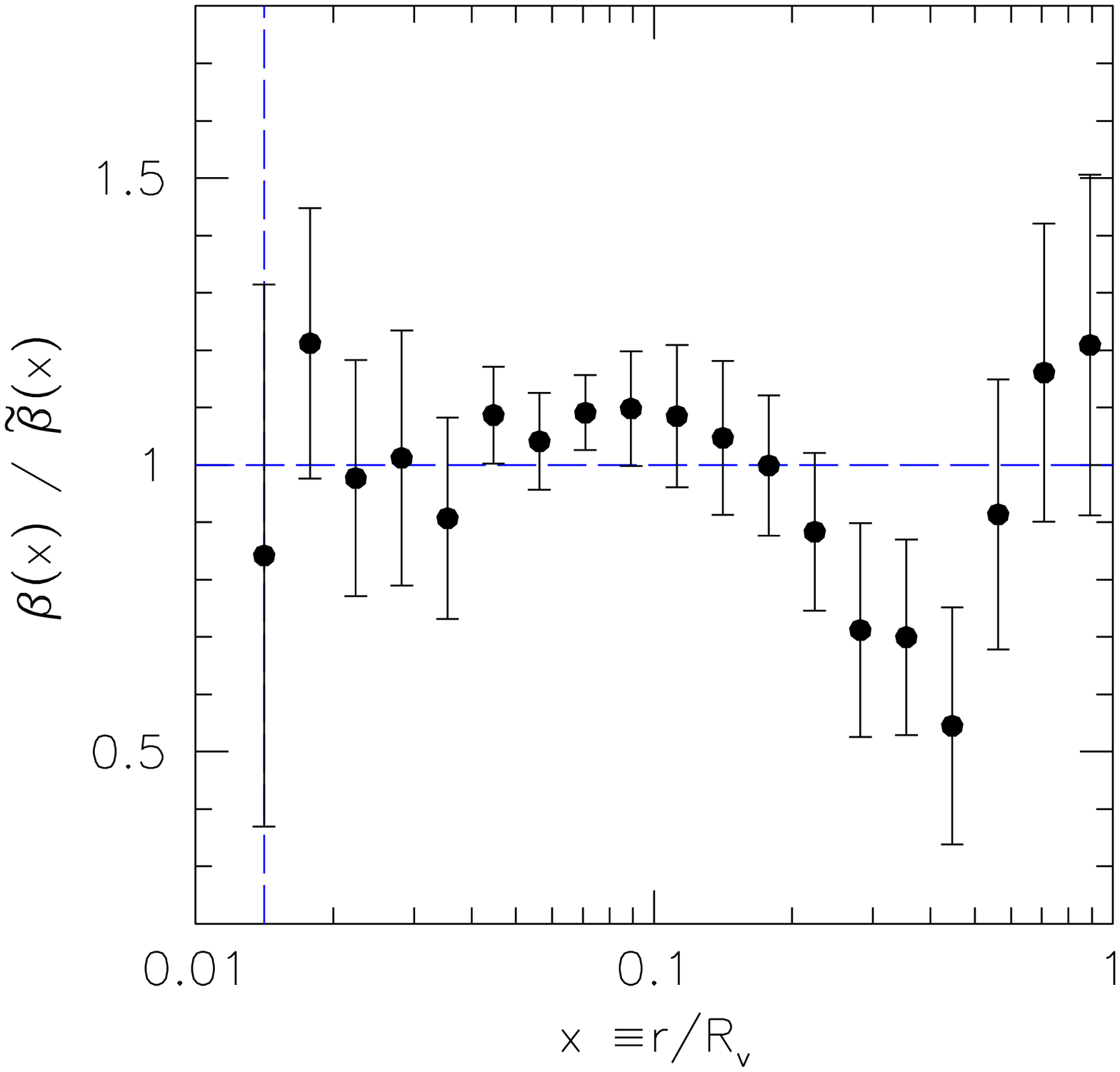}
\caption{Mean logarithmic residuals between the measured profiles and our
fitting formulae (equations~\ref{eq:sr}-\ref{eq:st}): radial velocity
dispersion $\sigma_r(x)$ (top panel), tangential velocity dispersion
$\sigma_t(x)$ (central panel) and velocity anisotropy $\beta(x)$
(bottom panel).  Residuals are plotted only for radii larger than the
fiducial limit of our fit: $\log x= -1.85$, as set by the analysis of
Fig.~\ref{fig:phaserho}.}
\label{fig:resvel}
\end{figure}

In Fig.~\ref{fig:resvel} we show the mean logarithmic residuals
between the actual profiles and our fitting formulae.  Error bars
represent the error on the mean.  Notice that for radial and
tangential velocity dispersions the differences are very small and
within $1 \sigma$ of the mean at most radii. The maximum discrepancies
are found at large radii, but they are still below 5 per cent.  The
velocity anisotropy $\beta(x)$ is equally well fitted at radii $x <
0.25$, while differences can be as large as 50 per cent at
intermediate distances: $0.3\la x \la 0.5$; this is the range where
$\sigma_t(x)$ is also showing the largest deviations from our fit.
Given the non-trivial shape of the measured velocity anisotropy
profile (bottom panel of Fig.~\ref{fig:sigdm}), we decided not to try
a more complicated fit for the sake of simplicity.  As we shall see
below, the error introduced in the Jeans equation by such deviations
is not relevant.

\subsection{Density structure}
\label{sec:dm_den}

The density profile of dark matter haloes has been widely studied in
the last few years (e.g. Crone, Evrard \& Richstone 1994; Jing et
al. 1995; Navarro, Frenk \& White 1996, 1997; Tormen, Bouchet \& White
1997; Huss, Jain \& Steinmetz 1999).  There is now a general agreement
that a double power-law with a central cusp is a good description of
simulated haloes (e.g. NFW), even though the exact value of the inner
asymptotic slope is still debated (Moore et al. 1998; Jing \& Suto
2000; Jing 2000; Ghigna et al. 2000; Power et al. 2003), possibly due
to the presence of systematic and/or numerical effects.

While it is not the purpose of this paper to enter this discussion, we
would like to stress that, unlike the majority of previous studies,
the dark matter profiles modelled in this work come from simulations
where the effect of the hot ICM is properly taken into account: the
presence of hot gas, even if not gravitationally dominant, can very
well affect also the dark matter structure.

A model for the dark matter density profile is straightforward to
obtain using the fitting relations derived in the previous
subsections.  If we combine the coarse-grained phase-space density
profile $\overline{f}(x)$ (equation~\ref{eq:phspace}) with the radial
velocity dispersion profile $\tilde{\sigma}_r(x)$
(equation~\ref{eq:sr}) and solve for the density, we easily get the
following approximate fit for the dark matter spatial density distribution:
\be
\label{eq:bel}
\tilde{\rho}(x) \equiv \frac{\rho(x)}{\rho_{\rm b}}
= \frac{\rho_0}{x(x+x_{p_1})^{1.5}} \ ,
\ee
with $\rho_{\rm b}$ the mean background density of the universe; the
normalisation factor $\rho_0$ is given by
\be
\label{eq:normbel}
\rho_0=\frac{(1 - f_{\rm b}) \Delta_{\rm v}}{6 \,\left[(1+2x_{p_1})/
(1+x_{p_1})^{1/2}-2 x_{p_1}^{1/2} \right]}\ ,
\ee
where $\Delta_{\rm v}$ is the virial overdensity specified by the
cosmological model; the term $f_{\rm b} = 0.097$ is the average
baryonic fraction of our cluster sample, measured at the virial
radius; it is used to properly weigh the DM component in our simulated
clusters.

Notice that for simplicity we have rounded the asymptotic behaviour of
this fit to $x^{-1}$ and $x^{-2.5}$ at very small and very large
scales, respectively. The exact slopes coming from the combination of
the two fitting relations for $\overline f(x)$ and $\sigma_r(x)$ would
be slightly different, $x^{-1.05}$ and $x^{-2.49}$ respectively.
However, as we will see, such small differences will not affect our
global model.

\begin{figure}
\centering
\includegraphics[width=7.cm]{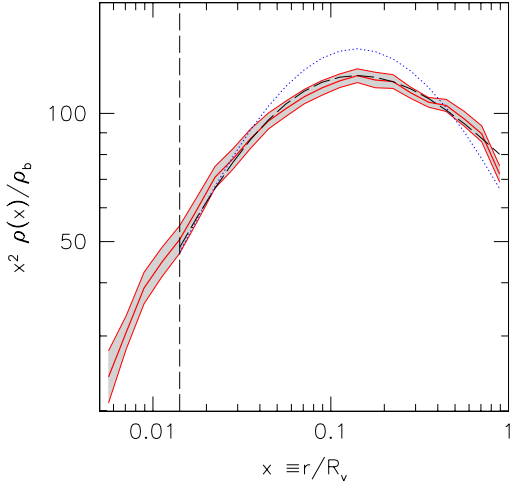}
\caption{Mean radial density profile for dark matter: $x^2 \rho(x)$ (solid
line) with error on the mean (shaded region).  The dashed and dotted
lines refer to our fit (equation~\ref{eq:bel}) and to the NFW fit
(equation~\ref{eq:nfw}), respectively.}
\label{fig:rhodm}
\end{figure}

The mean density profile, averaged over the whole sample of simulated
clusters, is shown in Fig.~\ref{fig:rhodm}.  As in the previous
figures, the shaded region shows the error of the mean, which is
relatively small, indicating that the different clusters have quite
consistent density profiles.  Our model (equation~\ref{eq:bel}) is
represented by the dashed line, and agrees very well with the data,
always falling inside the error region: only very close to the virial
radius there is a small discrepancy. This is a consequence of the
sharp change of slope observed in the mean profiles of both
$\overline{f}(x)$ and $\sigma_r(x)$, and cannot be improved without
adding significant complexity to our fitting formulae.

It is interesting to compare our relation to that originally proposed
by NFW, which is largely used in the literature to model the halo dark
matter distribution.  Its expression is the following:
\be
\frac{\rho_{\rm NFW}(x)}{\rho_{\rm b}}=\frac{\rho_{0,{\rm NFW}}}
{(x/x_s)(1+x/x_s)^2} \ .
\label{eq:nfw}
\ee
where $x_s$ is the scale parameter (inverse of the concentration
parameter: $c = 1/x_s$), and corresponds to our $x_p$, while
$\rho_{0,{\rm NFW}}$ is the normalisation factor or `characteristic
density':
\be
\label{eq:nfwnorm}
\rho_{0,{\rm NFW}}=\frac{(1 - f_{\rm b})\Delta_{\rm v} c^3}
{3[\ln(1+c)-c/(1+c)]}.
\ee
Since $\rho_{\rm b}$, $\Delta_{\rm v}$ and $f_{\rm b}$ are specified
by the cosmological model, for any given cosmology, $\rho_{0,{\rm
NFW}}$ is only a function of $x_s$, or of the concentration $c$.

\begin{figure}
 \centering
\includegraphics[width=7.cm]{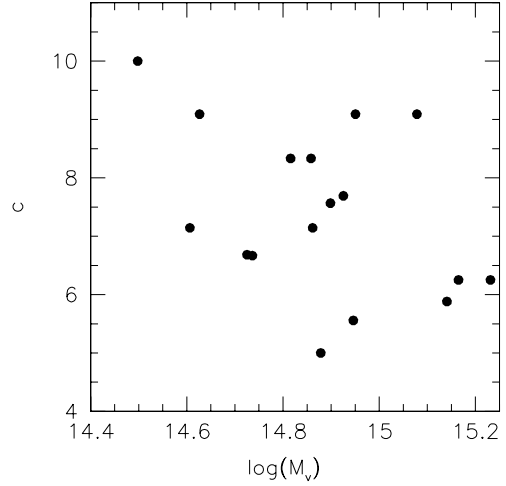}
\caption{Distribution of the concentration parameter $c$ versus 
the virial masses $M_{\rm v}$, in units of $h^{-1} M_{\odot}$.  Each
point refers to a different simulated galaxy cluster.}
\label{fig:conc}
\end{figure}

We found that the value of the NFW concentration parameter which best
fits our mean density profile is $c_{\rm NFW} \approx 6.8$.  This
value is slightly lower than the mean concentration over our cluster
sample, which is $\langle c\rangle = 7.13$.  However, the difference
is well within the scatter in the distribution of concentrations, as
shown in Fig.~\ref{fig:conc}, where the best fitting concentration for
each galaxy cluster is plotted versus the cluster virial mass.  The
figure also shows that there is no clear systematic trend with mass;
consequently it is reasonable to fit the mean profile as
representative of the full sample.  The resulting NFW profile is shown
in Fig.~\ref{fig:rhodm} as a dotted line.  We notice that for
distances $0.04\la x\la 0.4$ the NFW fit overestimates the mean
density profiles of the simulations.

\begin{figure}
\centering
\includegraphics[width=7.cm]{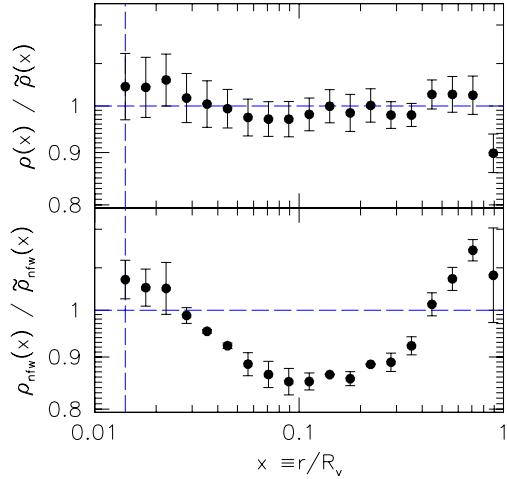}
\caption{Mean logarithmic residuals between the measured density 
profiles and the two fitting relations.  Top and bottom panels refer
to equations~(\ref{eq:bel}) and (\ref{eq:nfw}), respectively. Error
bars represent the error on the mean.}
\label{fig:resrho}
\end{figure}

This is more clearly visible in Fig.~\ref{fig:resrho}, where we plot
the mean logarithmic residuals between the cluster profiles and the
analytic fits (equations~\ref{eq:bel} and \ref{eq:nfw}); error bars
indicate the error on the mean.  The systematic trend shown by the
residuals to the NFW were already noticed in the literature
(e.g. Tormen et al. 1997), and indicate that the NFW fit overestimates
the actual profiles in the simulations by up to 15 per cent at
$x\approx 0.1$, with a rms residual of the mean of 12.1 per cent.
This behaviour is strongly reduced if the relation proposed in this
paper is used: in this case residuals are always within the error
bars, except at distances close to the virial radius, and have a rms
of 6.0 per cent.  This result is not necessarily at odd with the
literature; we wish to remind that the NFW fit was proposed to model
the density structure of isolated and relaxed dark matter haloes in
dark-matter-only simulations.  The fact that our clusters are not all
isolated and relaxed, and the presence of a hot gas component, can
both play a role in modifying the dark matter profile.  We will come
back to this issue in the discussion section.

\subsection{Mass Estimates}
\label{sec:dm_mass_est}

Given the previous relations, the cluster mass inside any given radius
can be obtained either by direct integration of the density profile,
or by applying the Jeans equation to the system.

Integration of the density profile (equation~\ref{eq:bel}) gives
immediately
\be
\tilde{M}(x) \equiv \frac{M(<x)}{M_{\rm v}} =\frac{6 \rho_0} {\Delta_{\rm
v}}\left[\frac{x+2x_p}{(x+x_p)^{1/2}} -2x_p^{1/2}\right]\ .
\label{eq:massabel}
\ee
The corresponding gravitational potential profile has also an analytic
expression:
\be
\tilde{\Phi}(x)=-16 \pi G \rho_0
\left[\frac{(x+x_p)^{1/2} - x^{1/2}}{x} \right].
\ee

For comparison, we also report the corresponding relations obtained by
integrating the NFW profile (equation~\ref{eq:nfw}):
\be
\frac{M_{\rm NFW}(< x)}{M_{\rm v}}=
(1 - f_{\rm b})\frac{ \ln(1+cx)-cx/(1+cx)}{\ln(1+c)-c/(1+c)}\
\label{eq:massaNFW}
\ee
for the mass, and
\be
\Phi_{\rm NFW}(x)=-\frac{4\pi G\rho_{\rm 0,{\rm NFW}}}{xc^3} 
\ln \left(1 + cx \right),
\label{eq:potNFW}
\ee
for the gravitational potential.

However, it is more interesting and useful to estimate the mass
profile through the Jeans equation which, for a spherical static
system, can be expressed as the ratio between the estimated and true
mass:
\be
M^E(< x) = -\frac{x R_{\rm v} \sigma_r^2(x)}{G}
\left[ \frac{d\ln \rho(x)}{d\ln x} +\frac{d\ln \sigma_r(x)}{d\ln x}
+ 2 \beta(x) \right].
\label{eq:jeans}
\ee
Tormen et al. (1997) and Thomas et al. (1998) have shown that
cluster-sized dark matter haloes are indeed reasonably modelled by
this equation.  Observed galaxy clusters have also been described by
such an equation (e.g. the CNOC cluster survey, Carlberg et al. 1997)
with results in agreement with those coming from simulations.  We
stress that we are not imposing the Jeans equation as a constraint to
our model: we are using it to test the self-consistency of our density
and velocity fitting formulae, assuming that the galaxy clusters obey
it.  Using the Jeans estimate allows in principle to also weigh the
different contributions to the total estimate, and to measure the
errors made when one or more terms in the equation are ignored for
lack of information on the system.

\begin{figure} \centering
\includegraphics[width=7.cm]{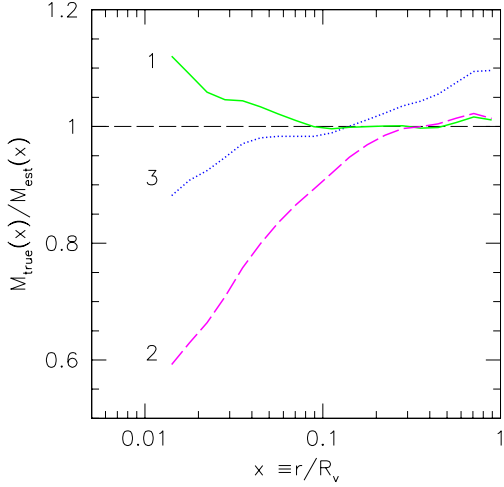}
\caption{Ratio between the true mass $M_{\rm true}(<x)$ and the one derived
from different estimators.  Different curves refer to $\tilde{M}(x)$
(equation~\ref{eq:massabel}; solid curve 1), $M_{\rm NFW}(<x)$ 
(equation~\ref{eq:massaNFW}; dashed curve 2); and $M^E(<x)$ 
(equation~\ref{eq:jeans}; dotted line 3).}
\label{fig:mass}
\end{figure}

The performance of the different mass estimators is shown in
Fig.~\ref{fig:mass}, where we plot the ratio between the actual mass
profile (i.e. the mean mass profile of our simulated clusters) and the
mass profile obtained by either integrating the density profile or by
inserting the analytic fits into the Jeans equation (\ref{eq:jeans}).
The profiles from simulations were multiplied by a factor $(1 +
\epsilon^2/x^2)^{3/2}$, where $\epsilon$ is the Plummer equivalent of
the actual spline softening, to deconvolve the effect of gravitational
softening and thus provide a more accurate comparison at small radii.
Curves labeled 1 and 2 refer to estimates based on the integrals of
the density models, $\tilde{M}(x)$ (equation~\ref{eq:massabel}) and
$M_{\rm NFW}(<x)$ (equation~\ref{eq:massaNFW}), respectively; curve 3
refers to the mass obtained from the Jeans equation (\ref{eq:jeans})
using the density and velocity profiles given in equations
~(\ref{eq:sr}), (\ref{eq:b}) and (\ref{eq:bel}).

Notice a very good agreement between the actual mass and that derived
by integrating our density profile (equation ~\ref{eq:bel}). The
maximum discrepancy is found at small radii ($\log x \approx -1.85$)
and is of the order of 10 per cent.  The analogous estimate from the
NFW density profiles shows larger discrepancies: roughly 40 per cent
at $\log x \approx -1.85$ and 10 per cent at $\log x \approx -1$. On
the other hand, both models accurately reproduce the mass profile
close to the virial radius, where the errors are of the order of 1-2
per cent.

The mass estimate obtained by the Jeans equation (curve 3) is also
very good; its error is always smaller than 10 per cent, suggesting
that the model we are proposing is indeed dynamically self-consistent.
The main discrepancy is found at distances close to the virial radius;
we have checked that this error originates mainly from the incorrect
slope of our density profile close to $R_{\rm v}$.

\section{Modelling Gas Profiles}

\subsection{Density structure}

\begin{figure} \centering
\includegraphics[width=7.cm]{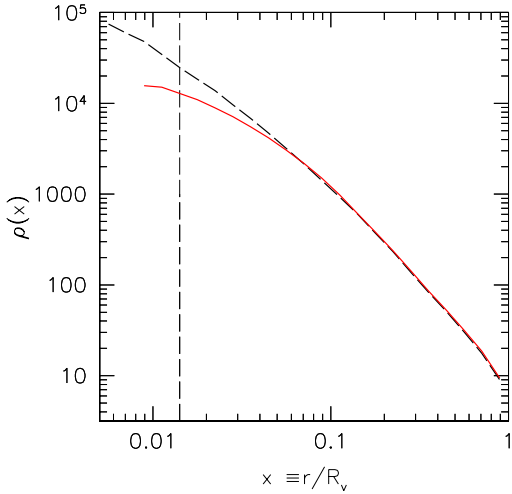}
\caption{Dark matter (dashed curve) and gas (solid curve) density radial
 profiles. The two curves are normalized to have the same value at the
 virial radius. The vertical line represents the limit we assumed for
 our model, at $\log x= -1.85$.}
\label{fig:rhogd}
\end{figure}

In analogy to what has been done in the previous section for the dark
matter profiles, in this section we will apply a similar analysis to
the radial profiles of the hot gas component.  Let us start from the
density. Fig.~\ref{fig:rhogd} compares the profiles of gas and dark
matter, normalised so that they match at the virial radius.  The plot
shows that the profiles are similar roughly for $x \ga 0.06$,
while the gas profile becomes flatter in the internal region. The
different internal slope of the two profiles agrees with previous
analyses, coming from simulations at smaller resolution (e.g. Navarro,
Frenk \& White 1995; Eke, Navarro \& Frenk 1998; Frenk et al. 1999;
Pearce et al. 2000; Lewis et al. 2000; Yoshikawa, Jing \& Suto 2000).
Recent analytical models of the ICM in galaxy clusters (Komatsu \&
Seljak 2001) also propose a gas density profile with a flatter inner
part.  In our case, however, the similarity extends to inner
radii than previously found.

The similarity of the DM and gas profiles in the outer part of
the cluster suggested us a first naive approach: that of fitting the
gas density profile with the same functional form used for the dark
matter, with a different concentration value, assuming of course a
different normalization based on the average baryon fraction.  The
best-fitting NFW profile for the mean gas density turned out to have a
concentration $c=4.3$.  However, the discrepancy (not shown) between
this model and the mean profile in the simulations resulted quite
large, the fit having too steep a slope in the inner part, where the
gas profiles becomes flatter than $x^{-1}$.

We therefore relaxed the constraint on the inner slope, and tried to
fit the gas density profile with a double power-law having the same
outer slope as the DM profile, but a different asymptotic behaviour at
small radii.  The best fitting result turned out to be the following
analytic expression:
\be
\tilde{\rho}(x)\equiv \frac{\rho(x)}{\rho_{\rm b}} = 
\frac{\rho_0}{(x+x_p)^{2.5}} ,
\label{eq:rhogas}
\ee
where $x_p=0.04$; the normalisation $\rho_0$ is given by
\be
\rho_0 = \frac{f_{\rm b} \Delta_{\rm v}}{3} 
\left[\frac{16x_p^3/3+40x_p^2/3 + 10
x_p+2}{(x_p+1)^{2.5}} -\frac{16 x_p^{1/2}}{3} \right]^{-1}.
\ee

\begin{figure} \centering
\includegraphics[width=7.cm]{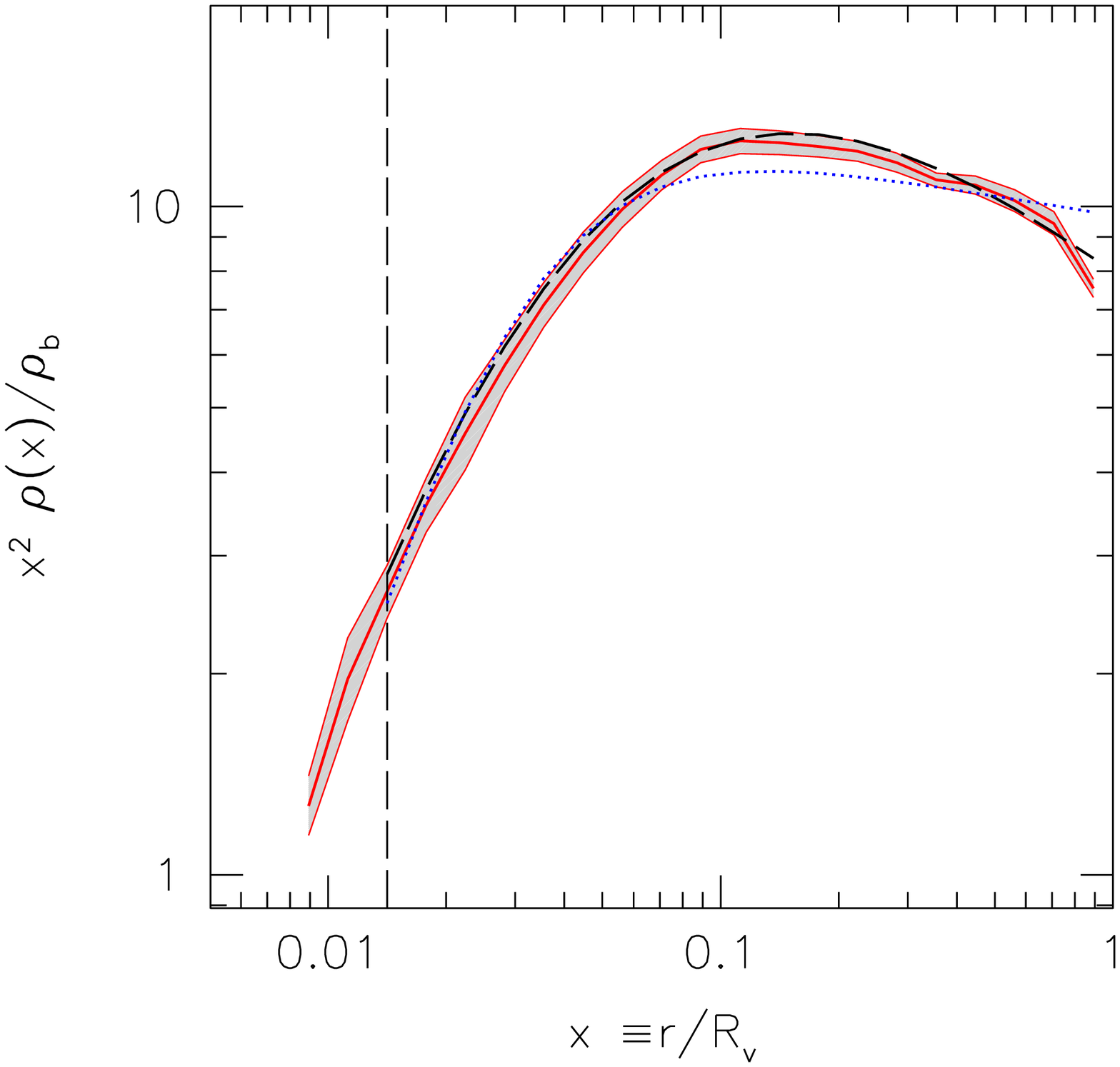}
\includegraphics[width=7.cm]{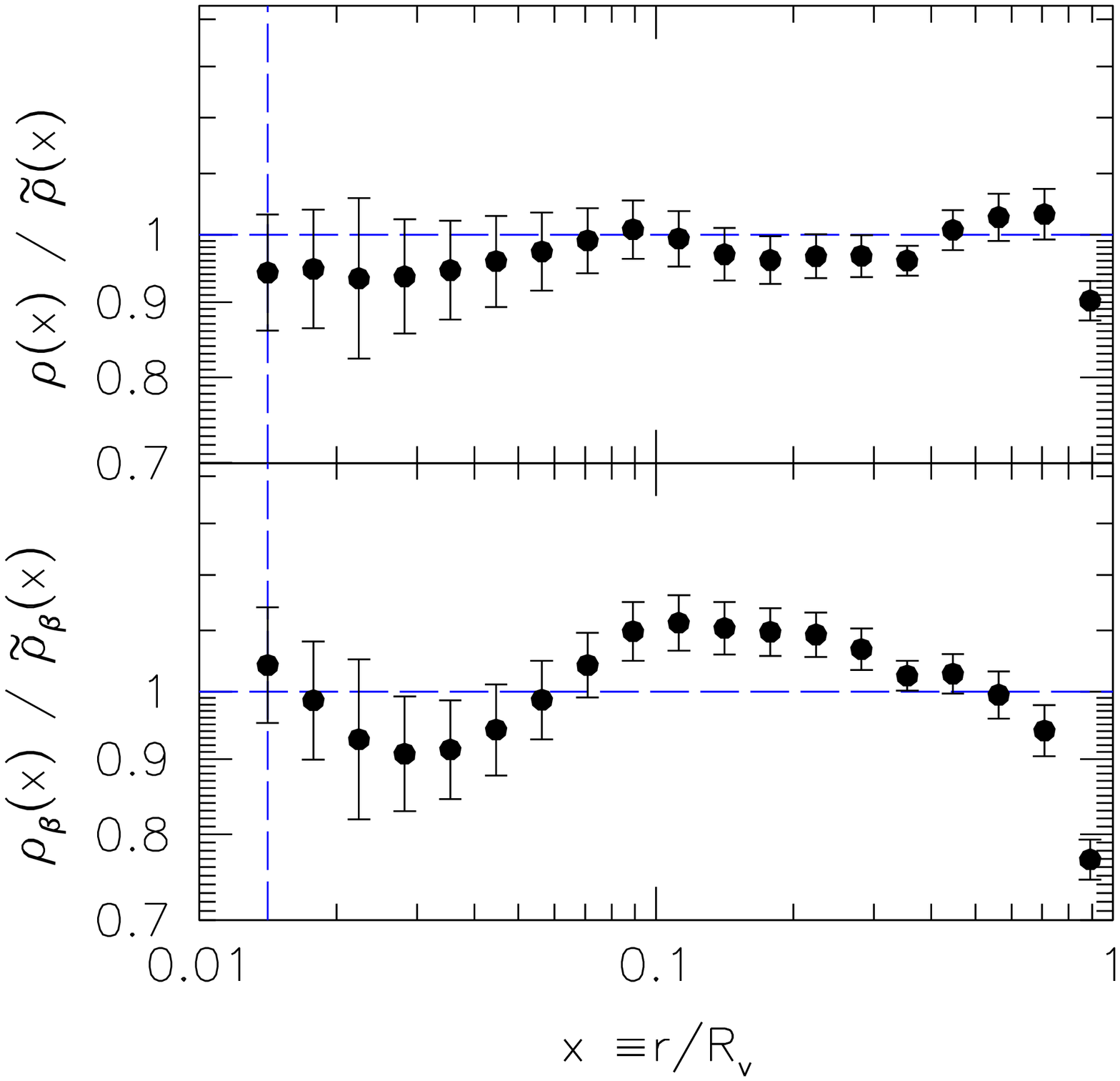}
\caption{Gas density profile.  Top panel: average radial profile $\rho x^2$ 
(solid curve) and its error (shaded region).  The dashed and dotted
curves represent our model (equation~\ref{eq:rhogas}) and a
$\beta$-model (equation~\ref{eq:betamodel}), respectively.  The
vertical line indicates the limit of our fit, taken at $\log x=
-1.85$.  Bottom panel: logarithmic residuals between the simulated gas
density profiles and the analytic models.  The upper and lower parts
of the plot refer to our model and to the $\beta$-model, respectively.
The points and error bars represent the mean ratio and their errors.}
\label{fig:rhogas}
\end{figure}

This fit tends to a constant value in the limit $x\to 0$ and decreases
as $\rho(x) \propto x^{-2.5}$ at large radii.  The upper panel of
Fig.~\ref{fig:rhogas} compares the proposed fit (dashed curve) to the
average profile for the gas density obtained from our simulated
clusters (solid curve): the agreement is always very good and the
fitting relation falls well inside the shaded region, which displays
the 1-$\sigma$ error of the mean. Notice that the results for our
simulated clusters are shown starting from the distance containing at
least 250 gas particles, while the fitting relations are computing for
$\log x>-1.85$, only.

A possible different approach is based on the so-called
{\it{$\beta$-model}} (Cavaliere \& Fusco-Femiano 1976, 1978) which is
the density profile most commonly used to describe the gas
distribution in observed galaxy clusters. Here the original underlying
assumption is that the gas density is proportional to the galaxy
density to a given power $\beta$.  Since the galaxy distribution was
assumed to follow a King profile, the resulting gas density profile is
given by
\be
\rho_\beta(x) = \frac{\rho_{0\beta}}{\left( x^2 + x_c^2 \right)^{3\beta/2}}\
,
\label{eq:betamodel}
\ee
where $x_c$ is the core radius (in units of the virial radius) and the
external slope is given by $3\beta$.  However, more recent analyses
have shown that the galaxy distribution traces the total mass
distribution, which is well described by a NFW profile (e.g. Carlberg
et al. 1997).  Moreover, equation~(\ref{eq:betamodel}) cannot fit - at
the same time - the central and external parts of the cluster
profiles, as traced by SZ and X-ray observations.  Finally, previous
numerical works have already shown that the $\beta$-model is unfit to
describe the results of (lower resolution) hydrodynamical simulations
(e.g. Navarro, Frenk \& White 1995; Bartelmann \& Steinmetz 1996).
Still, it is useful to test the $\beta$-profile against the
high-resolution simulations presented in this work.  We computed the
two parameters $x_c$ and $\beta$ by best fitting the mean profile of
the gas density, and obtained $x_c=10^{-1.53}$ and $\beta=0.7$.  The
corresponding $\beta$-profile is shown as a dotted curve in the upper
panel of Fig.~\ref{fig:rhogas}.  It is clear that there is not
agreement with the mean result of our simulations, being too flat at
large radii.  In order to quantify the goodness of fit for the two
models, in the lower panel of the same figure we show the logarithmic
residuals between the gas density of simulated clusters and our fit
(top) or the $\beta$-model (bottom).  While in the first case the
ratio is very close to unity for $x\la 0.8$, the $\beta$-model has a
10 per cent deviation for $0.1\la x \la 0.3$ and overestimates the
density by 25 per cent at the virial radius.

\subsection{Velocity structure}

\begin{figure}
\centering
\includegraphics[width=7.cm]{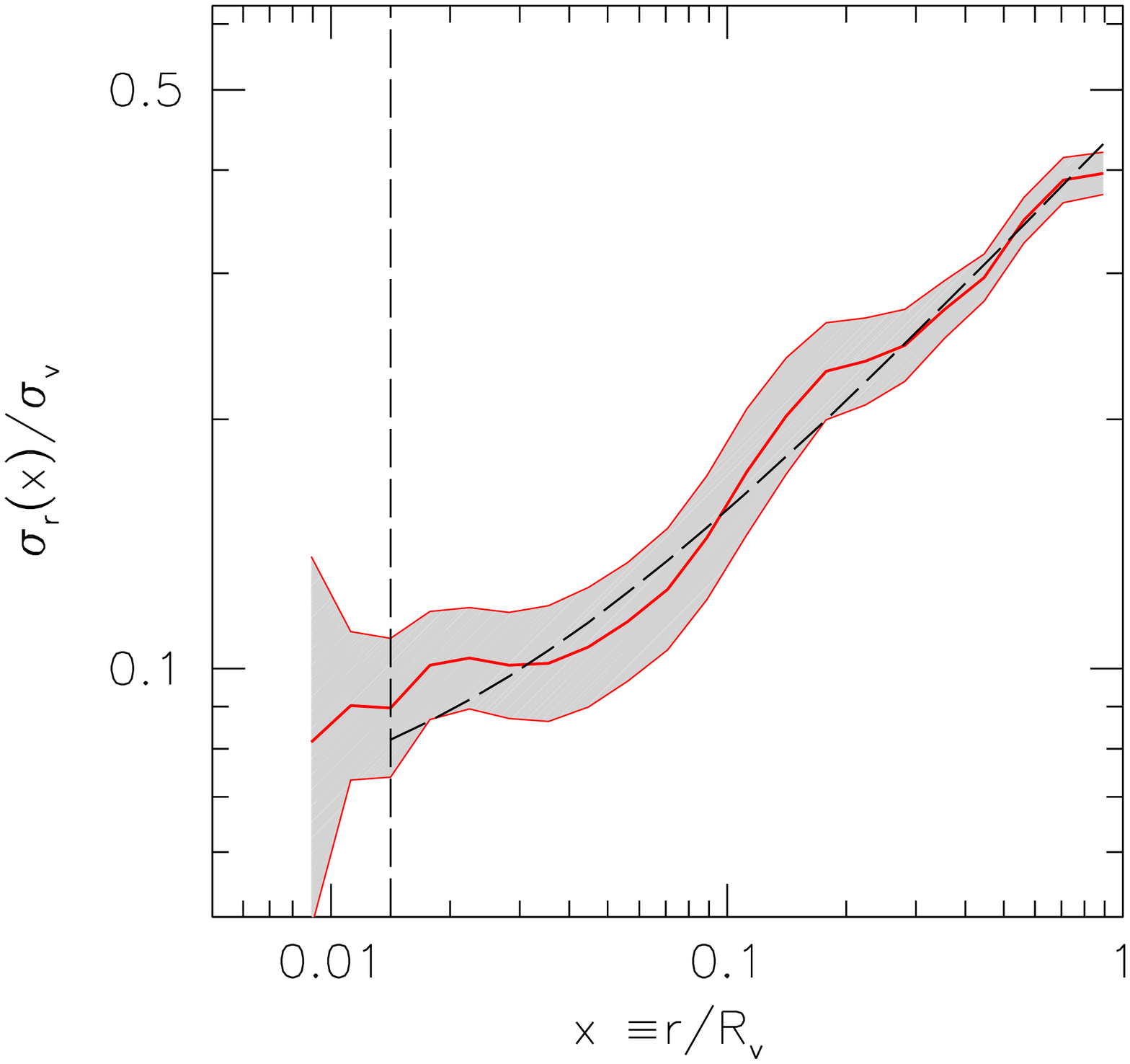}
\includegraphics[width=7.cm]{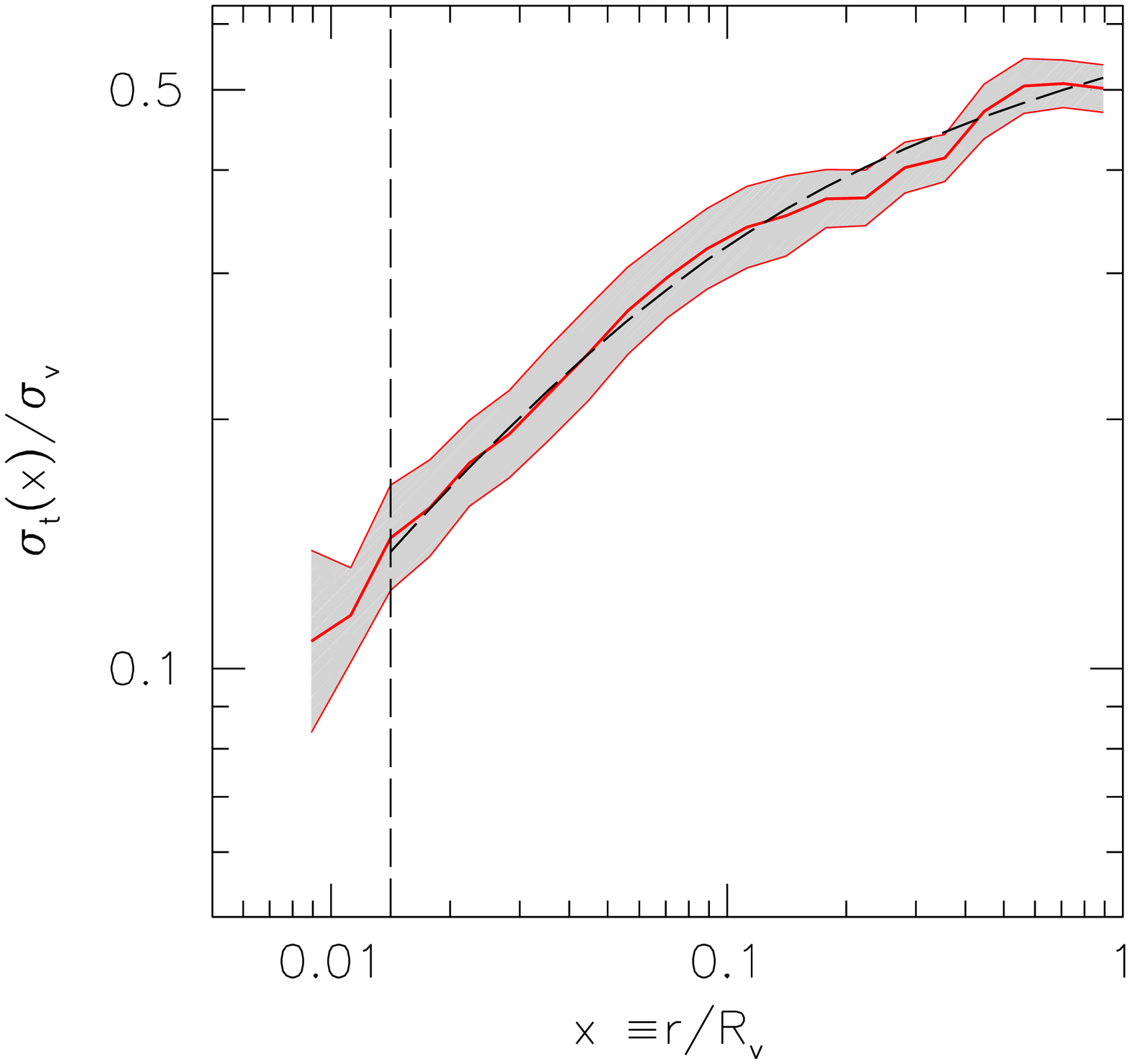}
\includegraphics[width=7.cm]{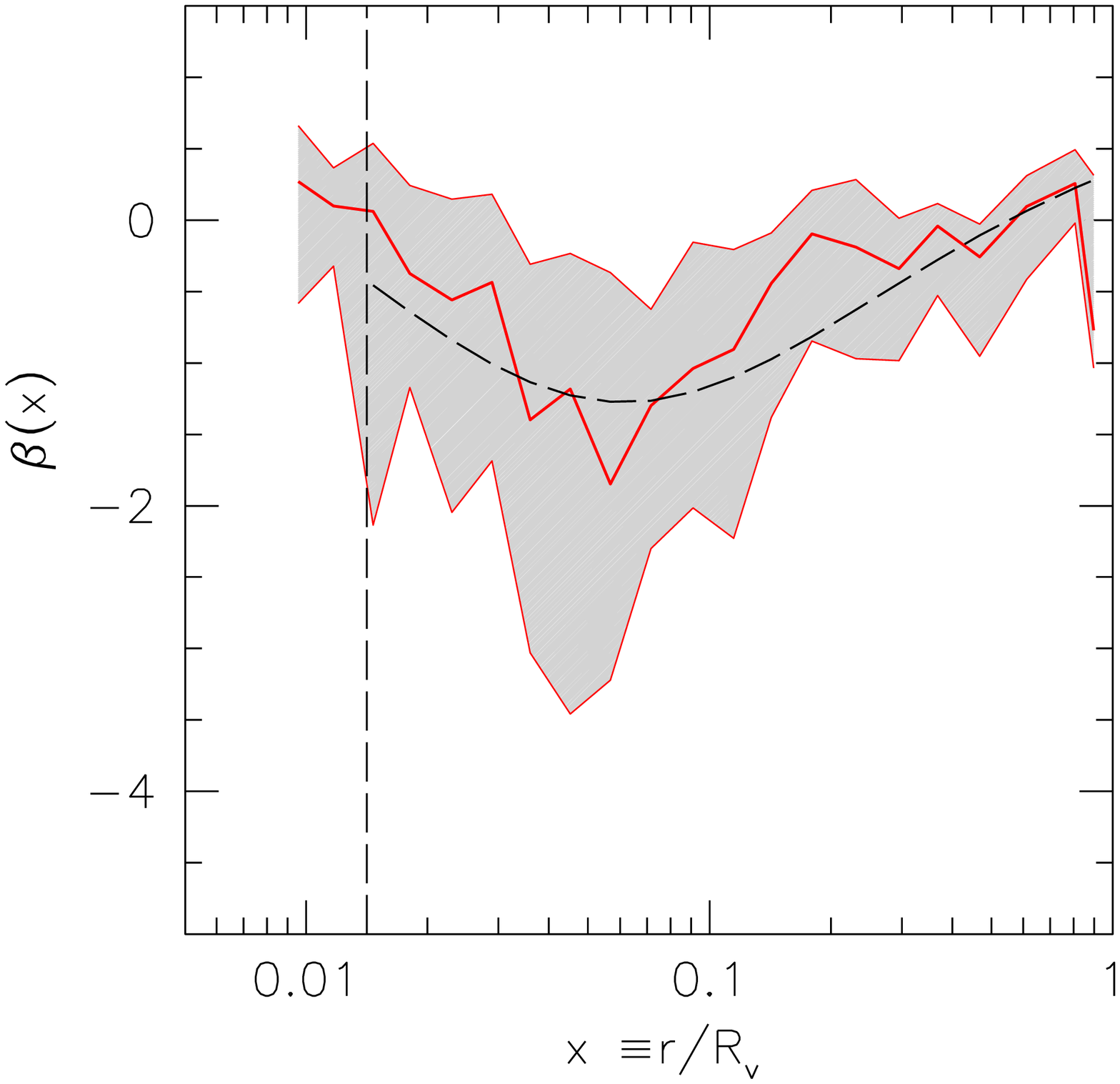}
\caption{
Radial velocity profiles for the gas: radial velocity dispersion
$\sigma_r(x)$ and tangential velocity dispersion $\sigma_t(x)$ (top
and central panels, respectively; the curves are normalized as in
Fig.~\ref{fig:sigdm}) and velocity anisotropy $\beta(x)$ (bottom
panel). In the upper and central panels solid lines and shaded regions
are the mean values obtained from our sample of simulated clusters and
their errors, respectively; only for $\beta$ we prefer to show the
median values and its errors. Dashed lines show our fitting
formulae (equations~\ref{eq:sigrgas}-\ref{eq:bgas}).  The vertical
dashed line represents the limit of our model, $\log x= -1.85$.}
\label{fig:siggas}
\end{figure}

We next consider the gas velocity dispersion profiles, normalized by
the virial velocity dispersion $\sigma_{\rm v}$ as we did for the dark
matter.  Results are shown in Fig.~\ref{fig:siggas}, where the solid
curve and the shaded region represent, as usual, the mean values from
the simulations, and the error on the mean.  Only for the profile of
the velocity anisotropy $\beta(x)$ we make use of the median instead
of the average: since $\beta(x)$ usually fluctuates around zero, an
average curve would be strongly affected by numerical fluctuations,
and would bias the mean curve away from zero.

We observed, as expected, that the gas velocity in
Fig.~\ref{fig:siggas} behaves very differently from the DM of
Fig.~\ref{fig:sigdm}, due to the different kind of forces acting on
the two components.  Both the radial and tangential gas velocity
dispersions are increasing functions of the distance $x$, even if
$\sigma_r(x)$ presents a central flattening.  The velocity anisotropy
$\beta(x)$ is marginally negative in the intermediate part of the
clusters ($x \approx 0.1$), denoting a slight predominance of
tangential motions.  Finally, the gas radial velocity (not shown) is
not completely zero, but slightly negative and larger at large radii,
suggesting a residual net infall.

Our best analytic fits to the curves for $\sigma_r$ and $\sigma_t$ are
also shown in Fig.~\ref{fig:siggas}, indicated by dashed lines; their
expressions are:
 \be
   \label{eq:sigrgas}
  \tilde{\sigma}_r(x) \equiv \sigma_r(x)/\sigma_{\rm v}= \sigma_{r_0}\ 
(x_{p_1}+x)^{0.5}
 \ee
 \bed
\textrm{where } \sigma_{r_0}=0.45 \textrm{ and } x_{p_1}=10^{-1.72};
\eed
\be
\label{eq:sigtgas}
   \tilde{\sigma}_t(x) \equiv \sigma_t(x)/\sigma_{\rm v}= \frac{\sigma_{t_0}
   x^{0.6}}{(x_{p_2}+x)^{0.5}}
\ee
\bed
\textrm{where\ } \sigma_{t_0}=0.54 \textrm{ and } x_{p_2}=10^{-1.1}.  
\eed

\begin{figure}
\centering
\includegraphics[width=7.cm]{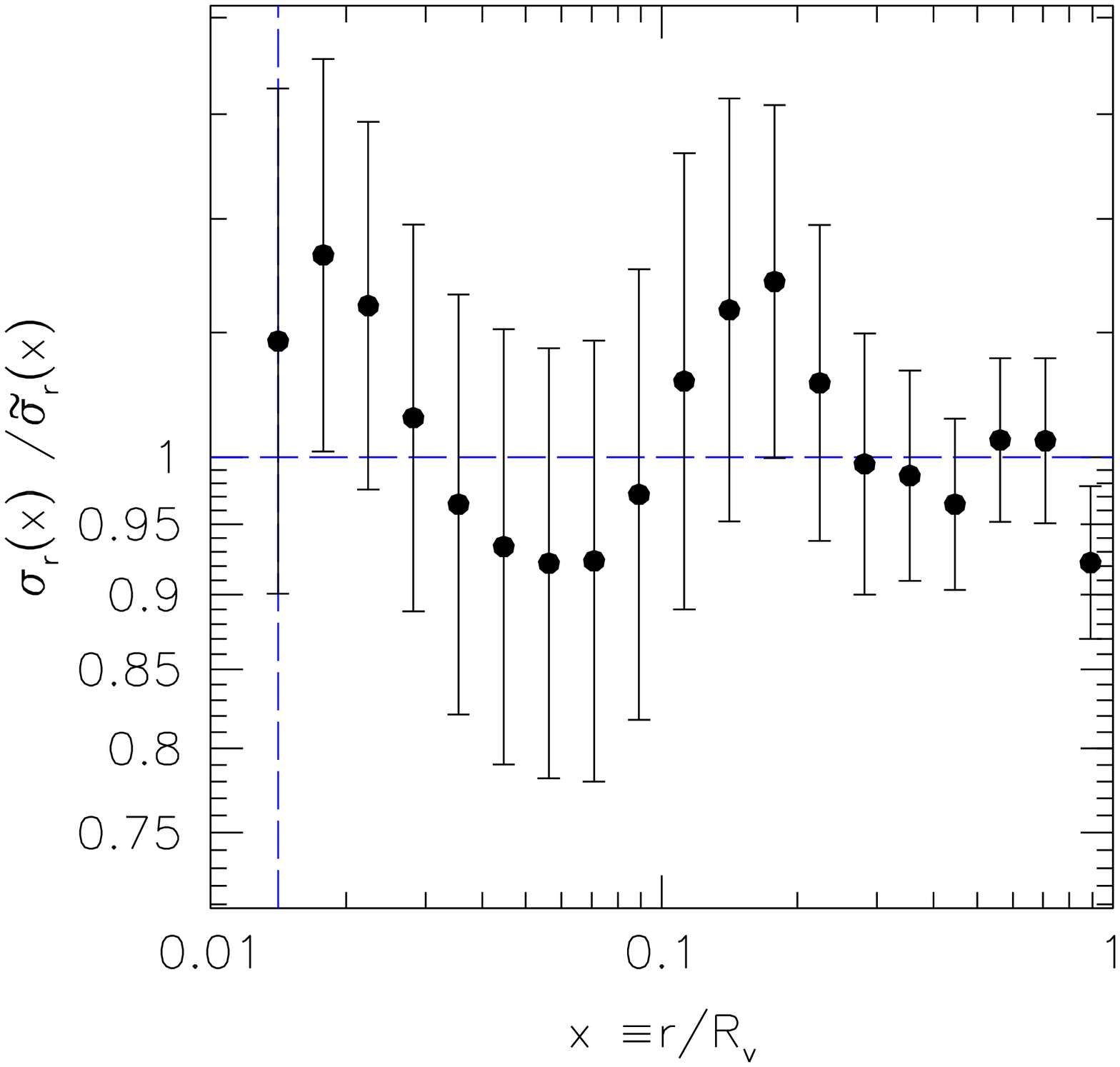}
\includegraphics[width=7.cm]{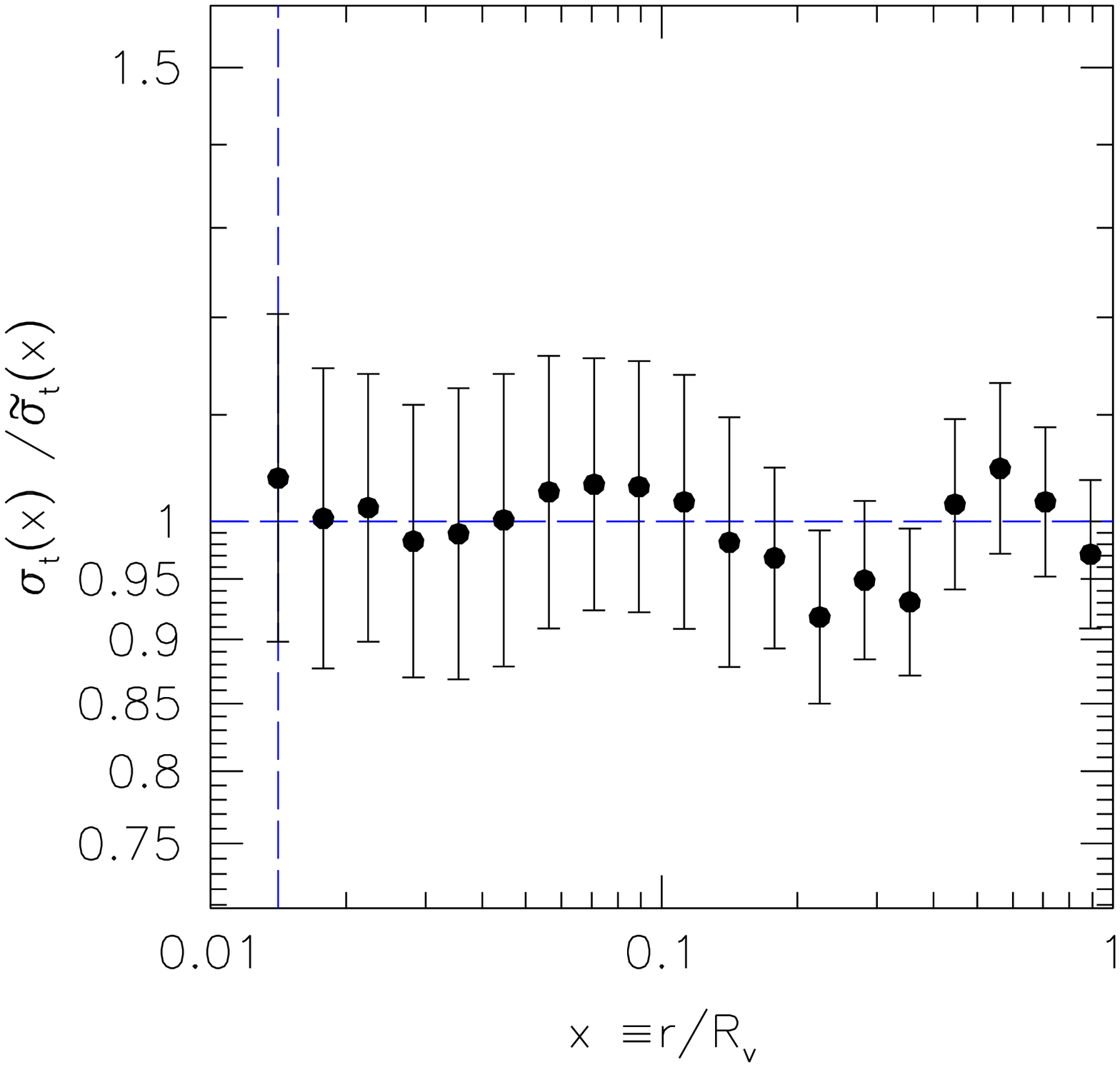}
\caption{Logarithmic residuals between the simulated gas velocity 
profiles and our models (equations~\ref{eq:sigrgas}-\ref{eq:sigtgas}).
Radial velocity dispersion $\sigma_r(x)$ and tangential velocity
dispersion $\sigma_t(x)$ are shown in the upper and bottom panels,
respectively.  The points and error bars represent the mean ratios and
the errors on the mean. The vertical dashed line represents the limit
of our model, $\log x= -1.85$.  }
\label{fig:resvelgas}
\end{figure}

The logarithmic residuals between the mean simulated profiles and our
models are plotted in Fig.~\ref{fig:resvelgas}, with mean values and
error indicated as usual.  The error bars are quite large, showing a
large variance in our cluster sample, but the results are always
compatible with unity, denoting a good agreement between simulated
clusters and models.
 
In order to write a model for the velocity anisotropy $\beta(x)$ we
decided to simply combine the two previous fits for the velocity
dispersions $\tilde{\sigma}_t(x)$ and $\tilde{\sigma}_r(x)$:
\be
\ \tilde{\beta}(x) \equiv  1 -
\frac{\tilde{\sigma}_t^2(x)}{2\tilde{\sigma}_r^2(x)} = 
1 - \frac{\sigma^2_{t_0} x^{1.2}}{2 \sigma^2_{r_0}(x_{p_1} + x)(x_{p_2} + x)}
\label{eq:bgas}
\ee
The resulting curve, displayed as dashed line in the corresponding
panel of Fig.~\ref{fig:siggas}, well reproduces the behaviour on the
median value, showing the self-consistency of our relations.


\subsection{Temperature structure}

The radial temperature profiles of galaxy clusters, as estimated from
X-ray observations, still show large uncertainties.  Only recently,
thanks to spectroscopic data coming from the Beppo-SAX and Chandra
satellites (see, e.g., Allen, Schmidt \& Fabian 2001; De Grandi \&
Molendi 2002; Ettori, De Grandi \& Molendi 2002a; Ettori et al. 2002b;
Johnstone et al. 2002), the complexity of the thermal structure inside
galaxy clusters has been revealed.  In particular, observations
suggest the presence of a central isothermal region followed by a
smooth decline towards the centre.  In order to reproduce this
feature, different recipes based on the inclusion of radiative cooling
and feedback heating have been unsuccessfully attempted in numerical
simulations (Lewis et al. 2000; Loken et al. 2002; Muanwong et
al. 2002; Valdarnini 2003; Tornatore et al. 2003; Borgani et
al. 2004).  We remind that these processes are not considered in our
non-radiative hydrodynamical simulations; however, they influence only
the very central part of galaxy clusters, and so in practice they do
not affect the model we propose here.  The behaviour of the
temperature profile is also observationally not well determined at
large radii and shows large variation from object to object. However,
there are strong indications that the value of the temperature at the
virial radius is approximately a factor of 2 smaller than at the
centre.  This fact in turn causes errors in the mass estimates
obtained using the equation of the hydrostatic equilibrium.  Lacking a
reliable measure of the radial temperature gradient in clusters based
on X-ray observations, it is highly useful to have some indications
from numerical simulations.  For these reasons we devoted great care
in modelling the mean temperature profile.

\begin{figure}
   \centering
\includegraphics[width=7.cm]{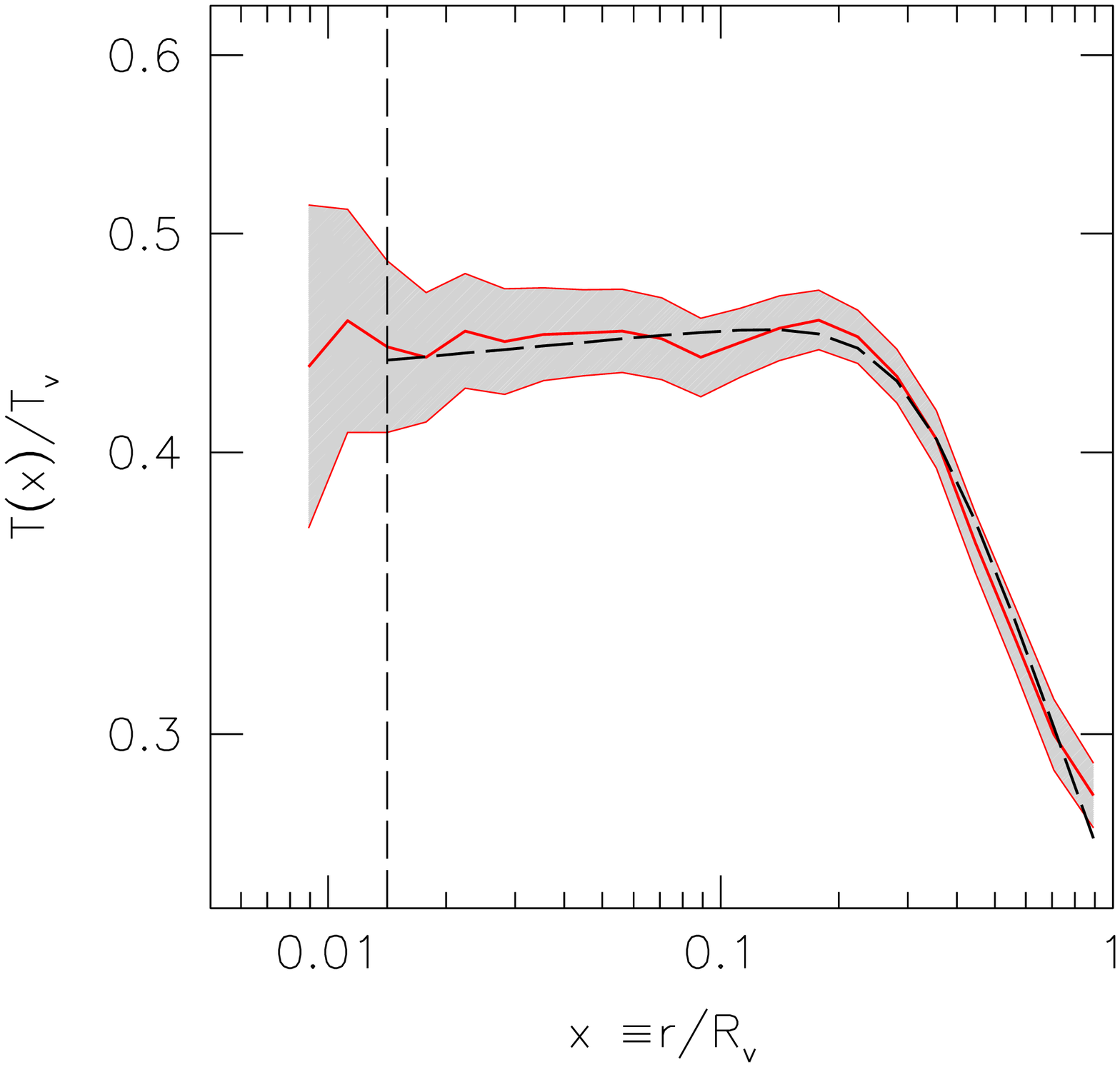}
\includegraphics[width=7.cm]{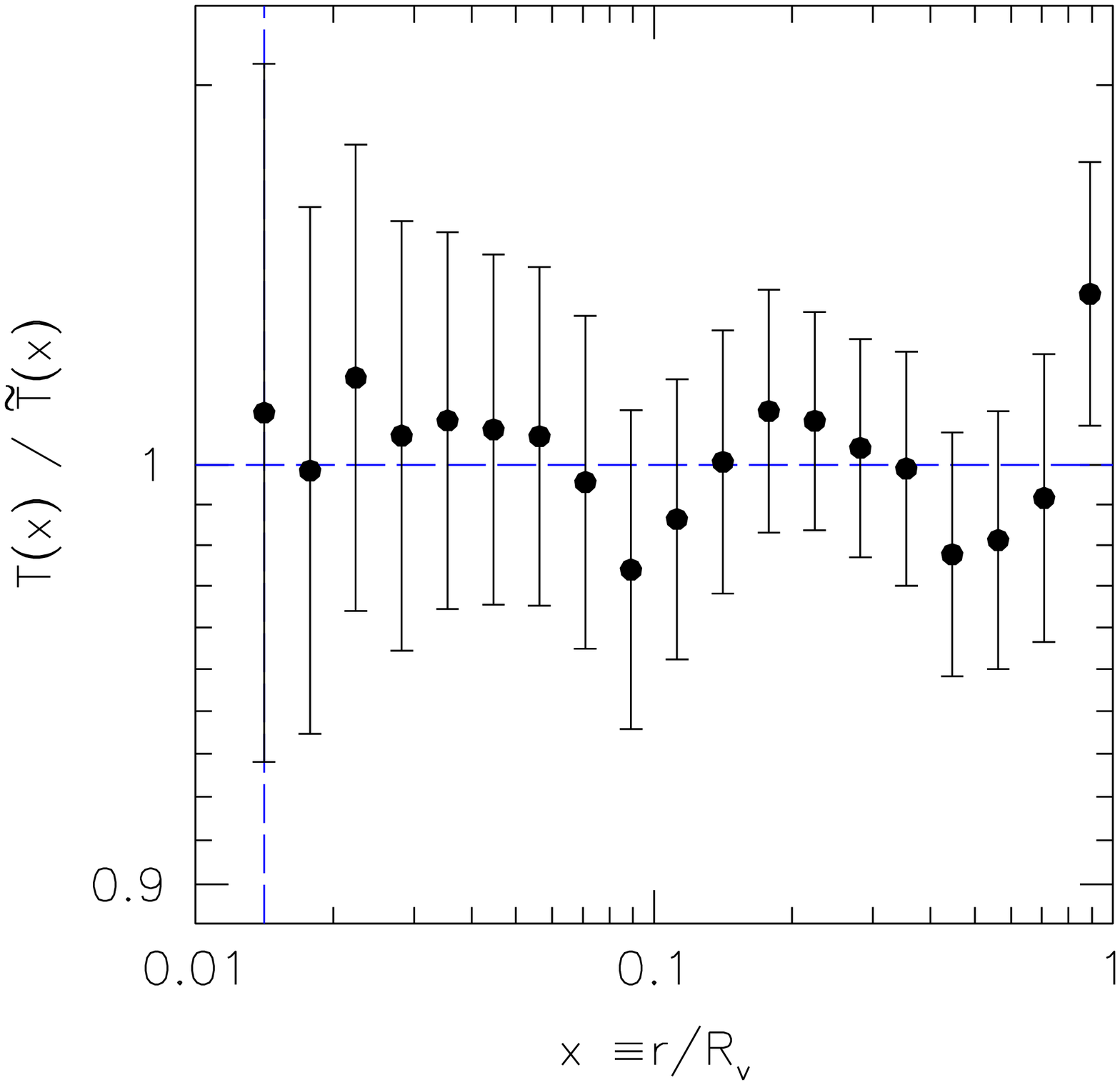}
 \caption{Gas temperature profile.  Top panel: mean radial profile
 $T(x)$ (normalised to the virial estimate $T_{\rm v}$; solid line)
 and its error (shaded region). The dashed line refers to our model
 (equation~\ref{eq:tem}).  The vertical line represents the limit of
 our model, $\log x= -1.85$. Bottom panel: logarithmic residuals
 between the simulated cluster temperature profile and the analytic
 model.  The points and error bars represent the mean ratio and the
 errors on the mean, respectively.  } \label{fig:tem}
\end{figure}

In the upper panel of Fig.~\ref{fig:tem} we show the average
temperature profile normalized by the estimate of the virial
temperature, $T_{\rm v}\equiv (G \mu m_p M_{\rm v})/ (k_{\rm b} R_{\rm
v})$.  In this relation $G$ is the gravity constant, $\mu=0.59$ is the
mean molecular weight, $m_p$ is the proton mass and $k_{\rm b}$ is the
Boltzmann constant.  The figure clearly shows a nearly isothermal
inner region, out to $x \approx 0.2$, with only a hint of positive
gradient.  At larger radii the temperature starts to decrease,
reaching at the virial radius a value about 60 per cent of the central
one.  The profile we find is in good agreement with previous numerical
works (Navarro et al. 1995; Eke et al. 1998; Bryan \& Norman 1998;
Frenk et al. 1999; Thomas et al. 2001; see, however, Loken et
al. 2002; AYMG03) and also with the temperature profile found from the
analysis of the Beppo-SAX data by De Grandi \& Molendi (2002).
However, we remark that the much higher numerical resolution of these
simulations makes us more confident in the result and allows us to
model the simulated profile with more accuracy.

The sudden temperature drop at $x \approx 0.2$ cannot be fit it by a
double power-law as those used so far. We model this sharp transition
using instead the following expression:
 \be
\tilde{T}(x)\equiv T(x)/T_{\rm v}=\frac{T_0 x^{0.016}}{(x^4 + x_p^4)^{0.13}},
 \label{eq:tem}
\ee
where $T_0 = 0.255$, $x_p = 10^{-0.51}$; the internal and external
logarithmic slopes are approximately $0$ and $-0.5$, respectively. The
bottom panel of Fig.~\ref{fig:tem} shows the logarithmic residuals
between the simulated profiles and the previous relation.  The mean
ratio is very close to unity, with no systematic trend, confirming the
accuracy of our fit.


\subsection{Mass Estimates}

In analogy to the dark matter, we can derive a first estimate of the
mass of hot gas inside any given radius by integrating the gas density
profile (equation~\ref{eq:rhogas}).  We obtain
\ba
\tilde{M}_{\rm ICM}(x) &\equiv& \frac{M_{\rm ICM}(<x)}{M_{\rm v}} = 
\frac{6 \rho_0}
{\Delta_{\rm v}} \nn &\times& \left[\frac{8x_p^3/3+20 x x_p^2/3 + 5
x^2 x_p + x^3}{(x_p+x)^{2.5}} -\frac{8 x_p^{1/2}}{3} \right]\ .
\label{eq:massgas}
\ea
The short-dashed curve labeled 6 in Fig.~\ref{eq:jeans} is the ratio
between the actual gas mass and the mass obtained from
equation~(\ref{eq:massgas}): the agreement between actual and modelled
mass profiles is excellent, with differences smaller than 5 per cent
at all radii $x >0.02$.  The expression in equation~(\ref{eq:massgas})
will be used below to discuss the baryonic fraction of our clusters.

We now have all the ingredients to discuss other mass estimates using
the gas profiles.  One of the standard assumption used to derive
cluster masses from X-ray data is that the system is in hydrostatic
equilibrium: the results of our high-resolution simulations can give
us some indications on the accuracy of this assumption, allowing us to
evaluate the size of the typical errors made in cluster mass
estimation when complete dynamical information are not available.

Assuming a spherical and static gravitational potential, the usual
equation for hydrostatic equilibrium can be formally obtained from the
Jeans equation (equation~\ref{eq:jeans}) by equating the gas internal
energy to the dark matter velocity dispersion $\sigma_{r,{\rm DM}}$,
\be
\label{eq:hydr1b}
  \frac{k_{\rm b} T}{\mu m_p} = \sigma_{r,{\rm DM}}^2
\ee
and by assuming an isotropic velocity field: $\beta(x) \equiv 0$.  One
then obtains the following mass estimate
\be
\label{eq:hydr1} 
M^E(<x) = -\frac{x R_{\rm v}  k_{\rm b} T(x)}
{G\mu m_p} \left[ \frac{d\textrm{ln}\rho(x)}{d \textrm{ln} x}+
\frac{d \textrm{ln} T(x)}{d \textrm{ln} x} \right].
\ee
However, the hypotheses underlying equation~(\ref{eq:hydr1}) are too
restrictive for at least two reasons.  First, our simulations show
that the mean velocity anisotropy $\beta(x)$ is not zero inside the
virial radius (see Fig.~\ref{fig:siggas}). Second, the replacement of
the velocity dispersion with the temperature is not fully justified,
due to the residual bulk motions shown in the same figure.

In order to have a more complete description of the gas behaviour, it
is more convenient to start directly from the force equation, and to
keep all terms coming both from isotropic pressure and from
anisotropic velocity dispersion.  For a system in a spherical and
symmetric potential $\Phi(x) = G M(<x) /x $, this {\em gas dynamical
equilibrium} equation can be written as:
\be
   \frac{d\Phi(x)}{dx} = \frac{1}{\rho(x)} \frac{dP(x)}{dx} +
   \frac{1}{\rho(x)} \frac{d\left[ \rho(x) \sigma_r^2(x)\right]}{dx} + 
   2 \beta(x) \frac{\sigma_r^2(x)}{x} \ .
\ee
By substituting the expression for the pressure, $P(x)=\rho(x) k_{\rm
b} T(x)/\mu m_p$, it is possible to obtain a new mass estimator
$M^E(<x)\equiv M(<x) / M_{\rm v}$ for the total mass inside $x$ traced
by the gas distribution:
\ba
 M^E(<x) & = & - \frac{x R_{\rm v}  k_{\rm b} T(x)}{G \mu m_p} \left[ \frac{d
  \ln\rho(x)}{d \ln x}+\frac{d \ln T(x)}{d \ln x}\right] - \nn
 & & \frac{x \ R_{\rm v}\ \sigma_r^2(x)}{G}  \left[
\frac{d \ln \rho(x)}{d \ln x}+\frac{d
\ln \sigma_r^2(x)}{d \ln x} +2\beta(x) \right] .
   \label{eq:hydr2} 
\ea 
We then used this complete equation, together with some
simplifications of it, to estimate the cluster mass inside $x$ using
the radial profiles we proposed for the various quantities.  In
particular we considered the following possibilities:
\begin{enumerate}
\item
a $\beta$-model isothermal sphere, giving:
\be
M^E_1(<x) = - \frac{x  R_{\rm v}  k_{\rm b} T(<x)}{G \mu m_p} 
\frac{d\ln \rho_\beta(x)}{d\ln x}\ ,
\label{eq:isobeta}
\ee
where $T(<x)$ is the actual mean temperature inside $x$ and
$\rho_\beta(x)$ is the $\beta$-model defined by
equation~(\ref{eq:betamodel});

\item
an isothermal sphere with density profile given by our best fit to the
simulations, equation~(\ref{eq:rhogas}), giving:
\be
M^E_2(<x) = - \frac{x R_{\rm v}  k_{\rm b} T(<x)}{G \mu m_p} 
\frac{d\ln \tilde{\rho}(x)}{d\ln x}\ ; 
\label{eq:isobel}
\ee

\item
the usual hydrostatic equilibrium equation~(\ref{eq:hydr1}) with a
$\beta$-model gas density profile (equation~\ref{eq:betamodel}) and
our best fitting relation for the temperature profile
(equation~\ref{eq:tem}):
\be
M^E_3(<x)= -x M_{\rm v} \tilde{T}(x)\left[ \frac{d\ln \rho_\beta(x)}{d\ln x}
  +\frac{d\ln \tilde{T}(x)}{d\ln x} \right] \ ; 
\label{eq:eqhydrost_beta}
\ee

\item
the same equation of hydrostatic equilibrium (\ref{eq:hydr1}), where
the density and temperature profiles are our best fit
equations~(\ref{eq:rhogas}) and (\ref{eq:tem}):
\be
M^E_4(<x)= -x M_{\rm v} \tilde{T}(x)\left[ \frac{d\ln \tilde{\rho}(x)}{d\ln x}
  +\frac{d\ln \tilde{T}(x)}{d\ln x} \right]\ ;
\label{eq:eqhydrost}
\ee

\item
the more general model described by equation~(\ref{eq:hydr2}):
\ba
M^E_5(<x)& = & -x M_{\rm v} \tilde{T}(x) \left[ \frac{d \ln \tilde{\rho}(x)}
               {d \ln x}+\frac{d \ln \tilde{T}(x)}{d \ln x} \right] \nn
         &   &  + x M_{\rm v} \tilde{\sigma}_r^2(x) \left[ \frac{d \ln
                \tilde{\rho}(x)}{d \ln x} +\frac{d \ln
                \tilde{\sigma}_r^2(x)}{d \ln x} + 2\tilde{\beta}(x) 
\right], 
\label{eq:eqcompl}
\ea
where all the profiles are given by our best fits to the simulations,
equations~(\ref{eq:rhogas}), (\ref{eq:tem}), (\ref{eq:sigrgas}) and
(\ref{eq:bgas});

\end{enumerate}

 \begin{figure} \centering
\includegraphics[width=7.cm]{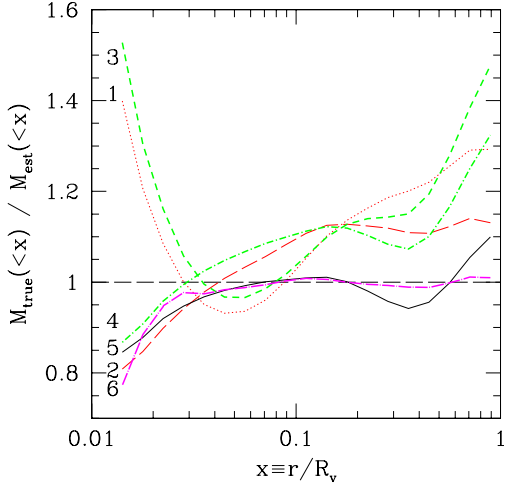}
\caption{
Ratio between the true mass $M_{\rm true}$ and the mass derived from
different estimators $M_{\rm est}$ presented in the text.  The curves
refer to $M^E_1$ (equation~\ref{eq:isobeta}: dotted line), $M^E_2$
(equation~\ref{eq:isobel}: long-dashed line), $M^E_3$
(equation~\ref{eq:eqhydrost}: short-dashed line), $M^E_4$
(equation~\ref{eq:eqhydrost_beta}: dotted-short-dashed line), $M^E_5$
(equation~\ref{eq:eqcompl}: solid line).  Curve labeled 6 represents
the ratio between the actual gas mass and the mass obtained from
equation~(\ref{eq:massgas}); dotted-long-dashed line. }
\label{fig:hydr}
\end{figure}

As done for the dark matter component in Fig.~\ref{fig:mass}, we
discuss the performance of the different mass estimators by plotting
in Fig.~\ref{fig:hydr} the ratio between the true mass profile,
i.e. the average mass profile of our simulated clusters, and the
previous analytic relations.  Curves labeled 1 and 2 refer to
estimates based on the isothermal sphere model ($M^E_1$ using
equation~\ref{eq:isobeta} and $M^E_2$ using equation~\ref{eq:isobel});
curves labeled 3 and 4 are for a standard hydrostatic equilibrium
model using for the temperature the new fit we propose and for the gas
density a $\beta$-model ($M^E_3$, equation~\ref{eq:eqhydrost_beta}),
or using for temperature and density our new fits ($M^E_4$,
equation~\ref{eq:eqhydrost}); curve labeled 5 is for the complete
`hydrodynamical equilibrium' model which includes the support from
residual gas motions, and uses all the analytical relations we derived
($M^E_5$, equation~\ref{eq:eqcompl}).

From the figure we notice that the $\beta$-models (curves 1 and 3)
underestimate the true cluster mass at most radii, the disagreement
being worst at large radii, where the isothermal $\beta$-model (curve
1) is off by up to 30 per cent, and the non-isothermal one (curve 3)
is off by up to 50 per cent. This is in agreement with previous
results in the literature (see e.g.  Muanwong et al. 2002).  Using our
density profile for the gas (curve 2 and 4) improves the estimate at
all radii of interest ($x > 0.2$), reducing the mass error at the
virial radius to 15 per cent (isothermal model, curve 2) and to
roughly 30 per cent (non-isothermal model, curve 4).  The complete
model (curve 5) is the most accurate at all radii, with an error of 5
to 10 per cent at $x > 0.2$.  Comparison of curves 4 and 5 shows that
the pressure contribution from bulk gas motions is significant at all
radii $x > 0.1$.

It is also interesting to see that the two isothermal estimators
(curves 1 and 2) on average fare better than the non-isothermal ones
(curves 3 and 4) at all $x > 0.2$.  The reason for this
counterintuitive result lies in the different temperature used in the
isothermal models (average temperature within each radius) compared to
the non-isothermal ones (local temperature at each radius).  By chance
this difference cancels part of the error made by ignoring the gas
motion in the outer cluster regions, and effectively makes isothermal
estimators more accurate than non-isothermal ones.  This is an
interesting result: if it were confirmed also when projection effects
and observational uncertainties are considered, it could lead to more
accurate mass estimations from X-ray observations.

We also note that the incomplete estimators (curves 1 to 4) generally
underestimate the actual cluster mass, but a simple isothermal model
with the appropriate gas density profile (curve 2) is still accurate
to the 15 per cent level.

\subsection{Baryonic Fraction}

A straightforward consequence of our model is an analytic model for
the baryon fraction radial profile, obtained by using the expressions
for the dark matter and gas masses (equations~\ref{eq:massabel} and
\ref{eq:massgas}):
\be
\label{eq:barfrac}
\tilde{f}_{\rm bar}(< x) \equiv \frac{ {\tilde M}_{\rm gas}(<x)}
              {\tilde{M}_{\rm gas}(<x) + \tilde{M}_{\rm DM}(<x)}\ .    
\ee
Notice that $\tilde{f}_{\rm bar}(x<1)\equiv f_{\rm b}=0.097$.

 \begin{figure} \centering
\includegraphics[height=7.cm]{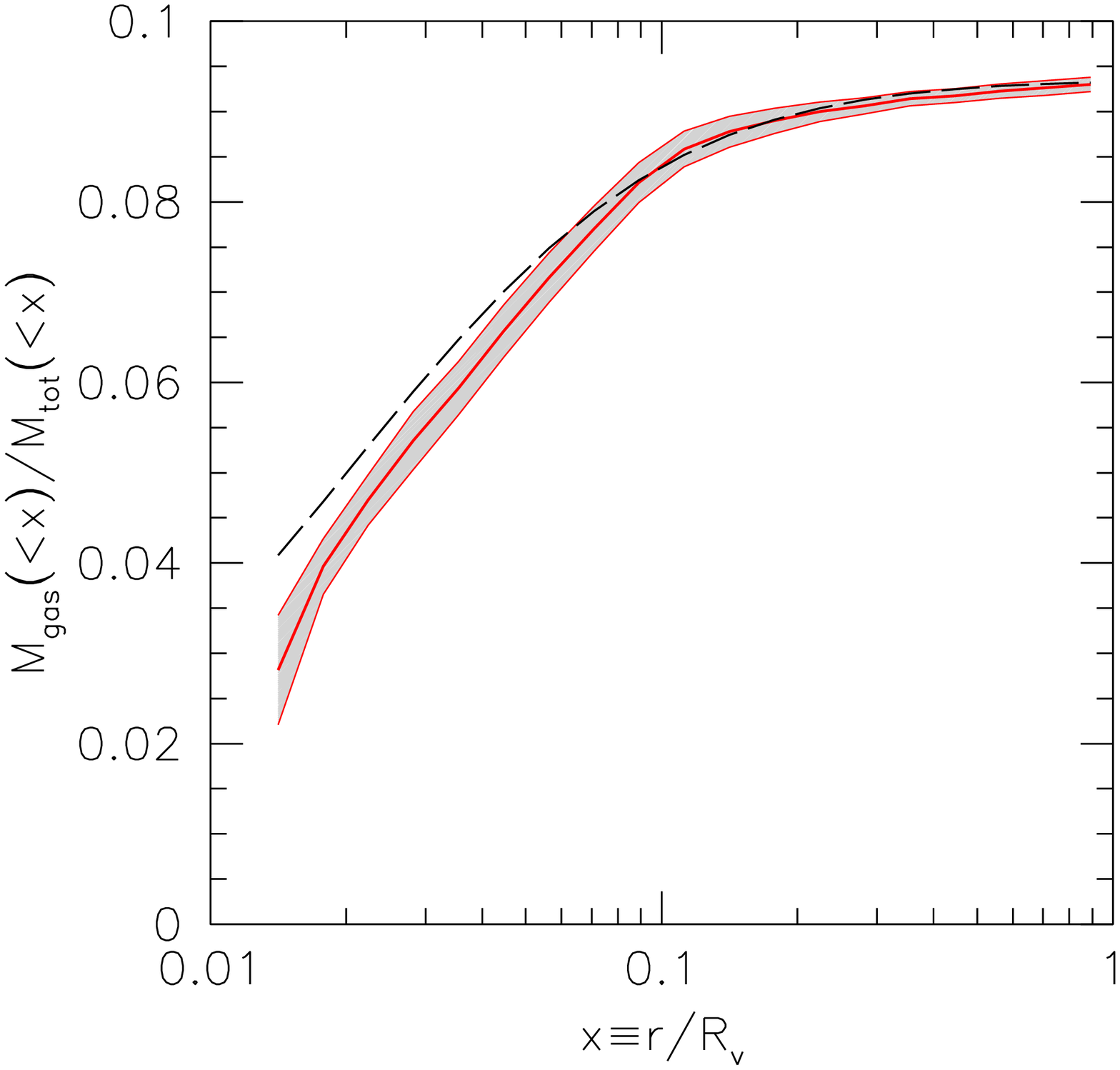}
\includegraphics[height=7.cm]{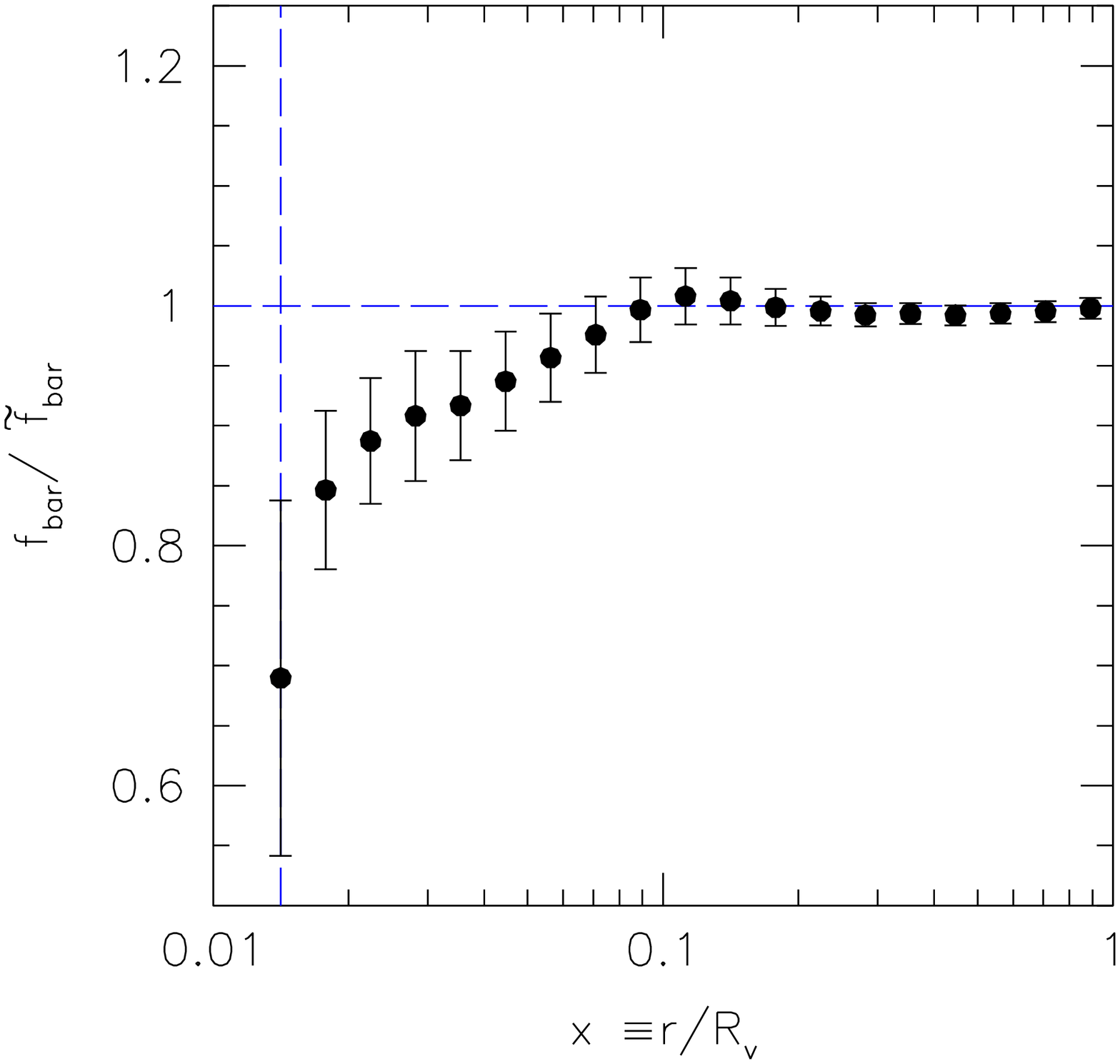}
\caption{
Baryonic fraction.  Top panel: radial profile of the baryonic
fraction.  Solid curve and shaded region indicate the mean and its
error for the cluster sample; the dashed curve is the baryonic
fraction predicted by integrating the analytic profiles proposed in
this paper (equation~\ref{eq:barfrac}). Bottom panel: logarithmic
residuals between the actual and predicted baryonic fractions.}
\label{fig:barfrac}
\end{figure}

The previous expression is compared to the actual baryon fraction in
Fig.~\ref{fig:barfrac}.  The analytic profiles predict a relation
(solid curve) that is practically indistinguishable from the actual
profile for $x > 0.1$, while at smaller radii deviations grows,
reaching a 15 per cent overestimate at $x = 0.02$.  However, this can
still be considered an excellent result of our analytic fits.


\section{Discussion}

\subsection{Effect of ICM on the density profile}

The results of Section \ref{sec:dm_den} on dark matter density
profiles indicate that the NFW fit has systematic residuals at
intermediate radii with respect to the actual mean dark matter profile
of our simulated haloes.  Although this was already shown by Tormen et
al. (1997) on a set of dark matter-only simulations, it is natural to
ask more generally whether and how much the presence of a hot gas
component affects the dark matter distribution.  All the SPH clusters
studied in this work were also simulated with dark matter only, using
otherwise the same initial conditions and identical parameters for the
evolutionary code: we can thus directly compare the density profiles
in the two instances.

\begin{figure} \centering
\includegraphics[height=7.cm]{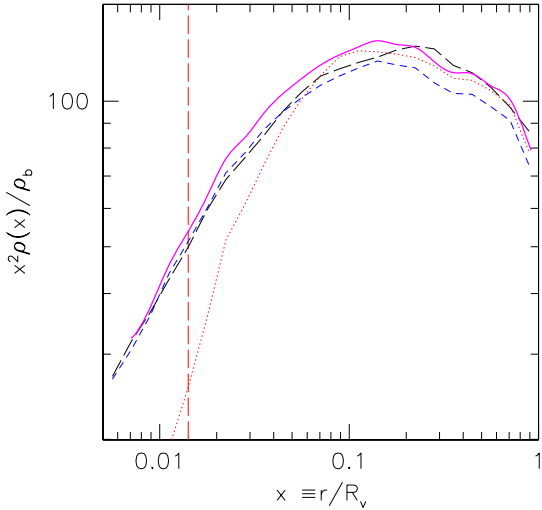}
\caption{
Comparison of density profiles in different simulations. The
long-dashed curve is the average dark matter profile in the pure
N-body runs; the short-dashed curve is the average dark matter profile
in SPH simulations; the dotted curve is the average gas profile in the
SPH simulations, moved up by a factor $f_{\rm b}^{-1}$, for clarity;
the solid curve is the total (dark matter plus gas) profile in SPH
simulations.}
\label{fig:compare}
\end{figure}

We would like to stress that the NFW profile was originally introduced
as a fit for the dark matter profile of isolated and relaxed dark
matter haloes in pure N-body simulations, and later it has been used
to fit the total profiles of observed clusters.  In both cases the
fitted profiles represent the total density distribution, not only the
dark matter one.  Therefore, we think it is here more appropriate to
compare the dark matter-only profiles of pure N-body clusters to the
total (i.e.  dark matter plus hot gas) profiles of our SPH clusters.
These are shown in Fig.~\ref{fig:compare}: the short- and long-dashed
curves represent the mean dark matter profiles in the SPH runs and in
the pure N-body companion runs, respectively, while the dotted curve
is the gas density (offset by a factor $f_{\rm b}^{-1}$ for clarity),
and the solid curve is the sum of dark matter and gas profiles in the
SPH runs.  Comparison of the two total profiles (long-dashed and solid
curves) shows that the effect of gas is to slightly concentrate the
total distribution.  An NFW fit on the long-dashed curve (total N-body
density) gives a mean concentration $c_{\rm DM}\approx 6.0$, while a
fit on the solid curve (total SPH density) gives a concentration
$c_{\rm SPH} \approx 6.5$.  Thus the effect of adding a hot gas
component shifts concentrations up by roughly 10 per cent, an amount
comparable to the baryonic fraction of our SPH runs.

\begin{figure} \centering
\includegraphics[height=7.cm]{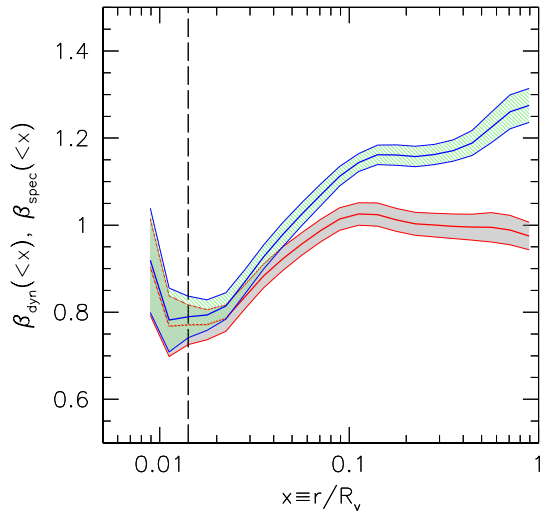}
\caption{
Cumulative profile of $\beta_{\rm dyn}$.  The curve and solid band
show the mean value (and its error) for equation~(\ref{eq:betadyn})
calculated in spheres of radius $x$.  The curve with hatched band
refers to $\beta_{\rm spec}$, i.e. it is obtained from the same
equation, but neglecting the gas bulk energy.}
\label{fig:betadyn}
\end{figure}

The explanation of this result might be found in the different energy
content of DM and gas.  As first pointed out by Navarro, Frenk \&
White (1995), during the accretion of matter onto a cluster, the
collisional nature of the ICM delays the gas infall compared to the
dark matter.  Due to this spatial lag, the gas feels a slightly
stronger gravitational field and so acquires extra energy at the
expenses of the dark component.  This fact was explicitly demonstrated
by Pearce, Thomas \& Couchman (1994) and by Navarro et al. (1995), but
can be also observed by comparing the energy content of the two
species through the ratio
\be
\label{eq:betadyn}
 \beta_{\rm dyn} \equiv \frac{\sigma_{\rm DM}^2}
{k_{\rm b} T/\mu m_p + \sigma_{\rm gas}^2} 
\ee
where $\sigma_{\rm DM}$ and $\sigma_{\rm gas}$ are the one-dimensional
velocity dispersions for the dark matter and gas, respectively, and
$T$ is the gas temperature.  This parameter differs from the usual
$\beta_{\rm spec}$ for the extra term $\sigma_{\rm gas}^2$, which
accounts for the gas bulk energy.  In Fig.~\ref{fig:betadyn} we show
the radial profiles of the average value of $\beta_{\rm dyn}$ in
spheres of given radius, $\beta_{\rm dyn}(<x)$. Solid curve and band
indicate mean over the cluster sample and its error.  This quantity
illustrates the energy ratio of dark matter and gas inside a sphere of
any radius $x$.  It clearly shows that, at radii larger than about $x
= 0.1$, there is energy equipartition between the two species.
Instead, at smaller radii, the gas has a larger energy content than
its collisionless companion.  The same radius is also close to where
the similarity of the gas and dark matter density profiles breaks
down, as shown in Fig.~\ref{fig:rhogd}: the extra energy acquired by
the gas makes it further expand, resulting in a flatter profile than
that of the DM.  For reference, the curve with hatched band is the
profile of $\beta_{\rm spec}(<x)$, and shows that neglecting the gas
bulk energy hides the existing energy equipartition at large radii.

The result of Fig.~\ref{fig:compare} seems to indicate that the energy
lost by the dark matter allows the halo to further compress and
concentrate compared to a pure collisionless collapse.  We may
speculate that, even if not large per se, this effect could lead to
observable differences in the properties of galaxy clusters at small
scales, e.g. with respect to gravitational lensing effects.

How well are these profiles fitted by the NFW formula?  The rms
residual of the mean profile for the pure N-body simulations is 16.7
per cent, comparable to previous findings (e.g. Tormen et al. 1997);
the residuals of the mean DM profile in SPH runs is 12.1 per cent; the
residual on the total (DM plus ICM) mean profile in the SPH runs is
13.3 per cent: therefore, besides changing the concentration, adding
the hot gas component slightly improves the NFW fit compared to the
pure N-body runs.  This seems to indicate that the departures of the
DM density from an NFW model are not due to the addition of gas (which
in fact slightly improves the fit), but rather to the fact that our
clusters were chosen with no specific criterion on their dynamical
status: they are not particularly relaxed, nor necessarily isolated.
Just for reference, if we try to fit the same profiles with our
relation for the dark matter (equation~\ref{eq:bel}) we obtain the
following rms residuals of the mean curve: 9.4 per cent, 6.0 per cent
and 4.6 per cent for DM in N-body runs, DM in SPH runs and total
density in SPH runs, respectively.  Therefore, the fit
(equation~\ref{eq:bel}) constitutes an improvement over the NFW in all
cases.  

\subsection{Dependence on the dynamical and environmental status}

As already said, we selected our sample in a random way, with no
constrain on the dynamics or environment of the objects.  On the other
hand, the NFW model is known to apply better to clusters which are
relaxed and virialized.  It is therefore important to investigate how
the fitting relations we propose in this paper depend on the dynamical
and environmental status.

To this aim, we first created a subsample excluding the systems which
show at $z=0$ evidences of a major merging event.  Following Tormen et
al. (1997), we defined as {\em perturbed} those clusters for which the
total mass, the dark matter velocity dispersion and the gas
temperature show a sudden significant increase (say at least 20
percent in mass) between the two last available outputs.  We found 5
such objects: excluding them we obtain a subsample of 12 relaxed
clusters, for which we computed the average profiles of all relevant
quantities. We found that the resulting average profiles are
statistically identical to those coming from the full sample; the only
exception is the radial velocity, which is much closer to zero at the
virial radius for the relaxed sample.  This is consistent with the
fact that selecting relaxed objects we tend to exclude systems with
significant infalling of matter.  More importantly for our
considerations, we found that all the fitting relations described in
the previous sections give again a proper description also for this
subsample.  In particular, this is true for the dark matter density
profile: this lets us conclude that the difference between the NFW and
our density profiles seems not to be due to the presence of merging
systems in the original sample.

\begin{figure} \centering
\includegraphics[height=7.cm]{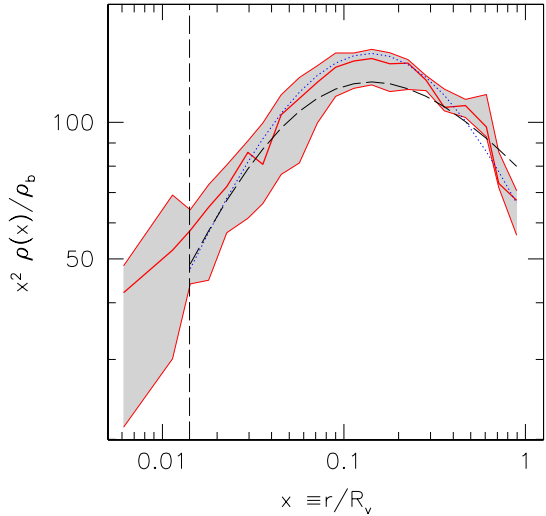}
\caption{Dark matter radial density profile for the subsample of 5
isolated clusters. The average and the corresponding error are shown by the 
solid line and shaded region.  Dashed and dotted lines refer to our
fit and to the NFW fit, respectively.}
\label{fig:rho_iso}
\end{figure}

A slightly different approach to select well-behaved objects is
considering the environment of each system: following Lanzoni (2000),
we defined as {\em isolated} those clusters (with virial mass $M_V$
and radius $R_V$) for which no companion with mass larger than 1/100
$M_V$ was found inside a sphere of radius 2 $R_V$.  In this way we
selected, from the original sample of 17, a subset of only 5 isolated
clusters (which is also a subset of the 12 relaxed ones).  The
resulting average profiles are similar to the previous ones, but now
the DM density profile (Fig.~\ref{fig:rho_iso}) shows a better
agreement with the NFW fit: in particular, the outer slope is somewhat
steeper, and the profile at intermediate region is more curved,
resulting in a reduction of the residuals. 

This fact suggests that the average density DM profile for our
sample deviates from the NFW fit mainly because of the inclusion of
non-isolated objects.

\subsection{Comparison with Ascasibar et al. (2003)}

The recent paper by AYMG03 has presented an analysis similar to ours,
applied to a sample of 15 simulated groups and poor clusters.  Their
approach is different in spirit, as they assume that the gas is in
hydrostatic equilibrium and that it follows a polytropic equation of
state, whereas we verify our model with the hydrostatic equilibrium
equation, and do not assume any equation of state.  Nevertheless, we
can compare some of their results with ours, once the different
definitions of cluster radii are taken into account.

In agreement with AYMG03 we also find that the $\beta$-model is not a
good fit to the gas profile of simulated clusters.  Instead, we do not
agree on the temperature profiles.  AYMG03 claim that a central
isothermal structure is a numerical artifact of standard SPH schemes
which do not implement entropy-conservation, and in fact we used the
public version of GADGET, which does not use the entropy-conservation
algorithm.  However, AYMG03 claim that differences between the two SPH
schemes are significant only at scales larger than $\approx 20 $ kpc
for their Cluster A, and indeed their Fig.~1 indicates that above
$\approx 0.03 R_{\rm v}$ the gas density and temperature profiles in
the two SPH implementations are statistically indistinguishable.  On
the other hand, our simulations exhibit isothermal cores out to
$x\approx 0.2$, a much larger scale.  Although we do not have a clear
explanation for this discrepancy, it could be that the disagreement is
at least enhanced by the different sample used (a sample of rich
clusters versus a sample of groups and poor clusters) and by the
different dynamical status (if the sample of AYMG03 is less
dynamically relaxed than ours, then the central temperature of their
systems would be on average perturbed and biased towards higher
temperatures).  We are currently creating and analysing a sample of
clusters at even higher resolution, simulated with the
entropy-conserving SPH scheme; the analysis of this new sample should
provide a definitive answer to this question.

As a consequence of having an isothermal core, we find that a single
polytrope is not a good description of the gas in our clusters.  While
at radii $x > 0.2$ the polytropic index $\gamma$ is close to $1.2$
(the value quoted by AYMG03), at smaller radii $\gamma$ drops to
unity.  AYMG03 plot their polytropic index only down to $0.2 R_{200}$,
roughly corresponding to $x=0.15$, so we cannot make a direct
comparison at small radii.

\section{Conclusions}

We have proposed a set of analytical fits which accurately describe
the mean radial properties of dark matter and diffuse gas in simulated
galaxy clusters.  Our relations have been calibrated on a sample of 17
high-resolution SPH clusters, chosen at random from a large parent
cosmological simulation. The model provides a self-consistent
description of these systems, and can be used to write accurate mass
estimates by mean of the Jeans equation for the dark matter
distribution and of a similar hydrodynamical equation for the diffuse
gas distribution.  For dark matter we propose a new model for the
density profile, obtained combining phase-space information and radial
velocity dispersion; the resulting profile describes our simulations
more accurately than the NFW model at $r < 0.7 R_{\rm v}$. This model
differs from the NFW one in that the asymptotic density slope at large
radii is $-2.5$ instead of $-3$.

More importantly, we introduce simple new fitting formulae for the DM
radial and tangential velocity dispersion profiles and for the
velocity anisotropy profile.  Radial and tangential velocity
dispersions profiles have a maximum value at intermediate radii and
decrease both towards the cluster centre and outskirts.  Velocity
anisotropy is an increasing function of radius: orbits are isotropic
in the centre and become increasingly radial orbits at larger radii.

These profiles are tested against the Jeans equation, and are shown to
give a self-consistent dynamical model, useful also for mass
estimates.  We find that the Jeans equation gives a realistic estimate
of the mass (with error typically less than 10 per cent).  We stress
that the focus of our analysis of the dark matter profile is not to
propose a replacement for the NFW fit, but to extend the model to all
the dark matter profiles relevant for studying the cluster internal
structure and dynamics.

We performed a similar study for the gas profiles, introducing new
analytic fits for gas density, temperature and velocity dispersions.
The average gas density profile of our sample is well described by a
double power-law, similar to the dark matter density at $r > 0.06
R_{\rm v}$, and approaching a constant density towards the centre.
The average temperature profile shows an isothermal core extending to
$r \approx 0.2 R_{\rm v}$, followed by a steep decrease that reaches a
factor two lower around the virial radius.

The gas velocity is not zero inside the clusters: residual motions
become smaller towards the centre - showing that the infalling gas
slows down as it sinks towards the cluster centre - and are isotropic
or slightly tangential.  Models for the gas velocity profiles are
required because the gas on average retains residual motions even
inside the virial radius, and this provides non-negligible anisotropic
pressure support to the equilibrium of the system.  For this reason we
generalized the hydrostatic equilibrium equation by adding velocity
terms analogous to those used in the Jeans equation for dark matter.
This `hydrodynamical equilibrium' equation is in fact a better
description of the system and provides a more accurate estimator of
the actual mass enclosed by any radius $x$.  However, if gas velocity
are neglected, a simple isothermal model fares better than a
non-isothermal one: if our gas density profile is adopted, then the
actual cluster mass is recovered to better than 15 per cent at all
radii of interest.

The dark matter and total density profiles in our SPH simulations
appear to be more centrally concentrated than those coming from
identical - but collisionless - simulations.  In terms of NFW halo
concentrations, these increase by an amount comparable to the baryonic
fraction.  Nevertheless, the addition of a hot gas component does not
degrade (but even improves) the goodness of fit of a NFW profile.
Therefore, the fact that our clusters are better fitted by the profile
proposed in equation~(\ref{eq:bel}) is not due to the presence of the
ICM. On the contrary we showed that it is due to the presence, in our
sample, of clusters in various environmental situations, as opposed to
having a sample of isolated systems.

The shape of the gas density profile at small radii is at least
partially explained by the gas expansion caused by the energy gain
from dark matter during the collapse.  Proper inclusion of the gas
bulk energy shows that the final energy budget is equally shared
between the two species at radii $r > 0.1 R_{\rm v}$, while at smaller
radii the gas is hotter than the dark matter.  This energy unbalance
is probably also the reason of the further global compression of the
system compared to a pure collisionless collapse.  It would be
interesting to investigate whether this effect is large enough to
produce observational signatures.

The analytical fits proposed in this work give an overall robust and
accurate description of the average dark matter and hot gas properties
of galaxy clusters.  They have immediate applications in cosmology, as
one can derive from them the radial dependence of many observables in
the X-ray, SZ and lensing domains.

Work is also in progress to extend this analysis to a sample of
cluster simulations where the gas is heated by non-gravitational
processes at high redshift, providing better agreement between the
X-ray properties of simulated and observed clusters.


\section*{Acknowledgments}
This work has been partially supported by Italian MIUR (Grant 2001,
prot. 2001028932, ``Clusters and groups of galaxies: the interplay of
dark and baryonic matter'') and ASI. GT and LM thank the Aspen Center
for Physics, where part of the paper was written up.  We are grateful
to Klaus Dolag, Gus Evrard, Carlos Frenk, Pasquale Mazzotta and
Massimo Meneghetti for useful discussions. We would like to thank the
referee, Peter Thomas, for his comments which improved the
presentation of our results.


\begin{thebibliography}{99}

\bibitem{allen} Allen S.W., Schmidt R.W., Fabian A.C., 2001, MNRAS, 328, L37

\bibitem{ascasibar} Ascasibar Y., Yespes G., Mueller V., Gottloeber S.,
2003, MNRAS, 346, 731 (AYMG03)

\bibitem{barnes} Barnes J.E., Hut P., 1986, Nat, 324, 446

\bibitem[\protect\citename{Bartelmann} 1998]{1998A&A...330....1B}
Bartelmann M., Huss A., Colberg J.M., Jenkins A., Pearce F.R., 1998, 
A\&A,  330, 1

\bibitem{bart} Bartelmann M., Steinmetz M., 1996, MNRAS, 283, 431

\bibitem{bert85} Bertschinger E., 1985, ApJS, 58, 39

\bibitem{borgani} Borgani S., et al., 2004,  MNRAS, in press, astro-ph/0310794

\bibitem{bria} Bryan G., Norman M.L., 1998, ApJ, 495, 80

\bibitem{carlberg} Carlberg R.G., et al., 1997, ApJ, 485, L13

\bibitem{cav} Cavaliere A., Fusco-Femiano R., 1976, A\&A, 49, 137

\bibitem{cav2} Cavaliere A., Fusco-Femiano R., 1978, A\&A, 70, 677
 
\bibitem{cole} Cole S., Lacey C., 1996, MNRAS, 281, 716

\bibitem[\protect\citename{Crone} 1994]{1994ApJ...434..402C} Crone M.M.,
Evrard A.E., Richstone D.O., 1994, ApJ, 434, 402

\bibitem{de grandi} De Grandi S., Molendi S., 2002, ApJ, 567, 1

\bibitem{eke96} Eke V.R., Cole S., Frenk C.S., 1996, MNRAS, 282, 263

\bibitem{eke} Eke V.R., Navarro J F., Frenk C.S., 1998, ApJ, 503, 569

\bibitem{ettori} Ettori S., De Grandi S., Molendi S., 2002a, A\&A, 391, 841

\bibitem{ettori2} Ettori S., Fabian A.C., Allen S.W., Johnstone R.M., 
2002b, MNRAS, 331, 635

\bibitem{frenk} Frenk C.S., et al., 1999, ApJ, 525, 554

\bibitem[\protect\citename{Ghigna} 2000]{2000ApJ...544..616G} Ghigna S., 
Moore B., Governato F., Lake G., Quinn T., Stadel J., 2000, ApJ,  544, 616 

\bibitem[\protect\citename{Hiotelis} 2002]{2002A&A...382...84H} Hiotelis 
N., 2002a, A\&A,  382, 84

\bibitem[\protect\citename{Hiotelis} 2002]{2002NewA....7..531H} Hiotelis
N., 2002b, NewA,  7, 531

\bibitem[\protect\citename{Huss} 1999]{1999ApJ...517...64H} Huss A., Jain 
B., Steinmetz M., 1999, ApJ,  517, 64 

\bibitem{jenkins} Jenkins A., Frenk C.S., White S.D.M., Colberg J.M.,
Cole S., Evrard A.E., Couchman H.M.P., Yoshida N., 2001, MNRAS, 321, 372

\bibitem{jing} Jing Y.P., 2000, ApJ, 535, 30

\bibitem[\protect\citename{Jing} 1995]{1995MNRAS.276..417J} Jing Y.P., Mo
H.J., Borner G., Fang L.Z., 1995, MNRAS,  276, 417

\bibitem{jingsuto} Jing Y.P., Suto Y., 2000, ApJ, 529, L69

\bibitem{jingsuto02} Jing Y.P., Suto Y., 2002, ApJ, 574, 538

\bibitem{johnstone} Johnstone R.M., Allen S.W., Fabian A.C., Sanders J.S., 
2002, MNRAS, 336, 299

\bibitem[\protect\citename{Komatsu} 2001]{2001MNRAS.327.1353K} Komatsu E.,
Seljak U., 2001, MNRAS,  327, 1353 

\bibitem{lanzoni} Lanzoni B., 2000, Ph.D. thesis, Universit\`e de Paris 7

\bibitem{lee} Lee J., Suto Y., 2003, ApJ, 585, 151

\bibitem{lewis} Lewis G.F., Babul A., Katz N., Quinn T., Hernquist L.,
Weinberg D.H., 2000, ApJ, 536, 623

\bibitem{loken} Loken C., Norman M.L., Nelson E., Burns J., Bryan G.L., 
Motl P., 2002, ApJ, 579, 571

\bibitem[\protect\citename{Meneghetti} 2003]{2003MNRAS.340..105M}
Meneghetti M., Bartelmann M., Moscardini L., 2003, MNRAS, 340, 105

\bibitem{moore} Moore B., Governato F., Quinn T., Stadel J.,
Lake G., 1998, ApJ, 499, L5

\bibitem{muanwong} Muanwong O., Thomas P.A., Kay S.T., Pearce F.R., 
2002, MNRAS, 336, 527

\bibitem{nfw1} Navarro J.F., Frenk C.S., White S.D.M., 1995,
MNRAS, 275, 720

\bibitem{nfw2} Navarro J.F., Frenk C.S., White S D.M., 1996, ApJ, 462,
563

\bibitem[\protect\citename{Navarro} 1997]{1997ApJ...490..493N} Navarro
J.F., Frenk C.S., White S.D.M., 1997, ApJ,  490, 493 (NFW)

\bibitem{pearce94} Pearce F.R., Thomas P.A., Couchman H.M.P., 
1994, MNRAS, 268, 953

\bibitem{pearce} Pearce F.R., Thomas P.A., Couchman H.M.P., Edge A.C.,
2000, MNRAS, 317, 1029

\bibitem[\protect\citename{Power} 2003]{2003MNRAS.338...14P} Power C.,
Navarro J.F., Jenkins A., Frenk C.S., White S.D.M., Springel V., Stadel
J., Quinn T., 2003, MNRAS,  338, 14

\bibitem[\protect\citename{Sheth} 2001]{2001MNRAS.325.1288S} Sheth R.K.,
Hui L., Diaferio A., Scoccimarro R., 2001, MNRAS,  325, 1288

\bibitem{spring01} Springel V., White S.D.M., Tormen G., Kauffmann G., 2001a,
MNRAS, 328, 726

\bibitem{spring} Springel V., Yoshida N., White S.D.M., 2001b,
NewA, 6, 79

\bibitem[\protect\citename{Suto} 1998]{1998ApJ...509..544S} Suto Y., Sasaki
S., Makino N., 1998, ApJ,  509, 544

\bibitem[\protect\citename{Taylor} 2001]{2001ApJ...563..483T} Taylor J.E.,
Navarro J.F., 2001, ApJ,  563, 483 

\bibitem{thomas} Thomas P.A., Muanwong O., Pearce F.R., Couchman H.M.P.,
Edge A.C., Jenkins A., Onuora L., 2001, MNRAS, 324, 450

\bibitem{thomas98} Thomas P.A. et al., 1998, MNRAS, 296, 1061

\bibitem{tormen2} Tormen G., Bouchet F.R., White S.D.M., 1997,
MNRAS, 286, 865

\bibitem{tormen2003} Tormen G., Moscardini L., Yoshida N., 2004, MNRAS,
in press, astro-ph/0304375

\bibitem{tornatore} Tornatore L., Borgani S., Springel V., Matteucci F., 
Menci N., Murante G., 2003, MNRAS, 342, 1025

\bibitem{valda} Valdarnini R., 2003, MNRAS, 339, 1117

\bibitem{yoshida} Yoshida N., Sheth R.K. Diaferio A., 2001, MNRAS, 328, 669

\bibitem{yokishawa} Yoshikawa K., Jing Y.P., Suto Y., 2000, ApJ, 535, 593
\end{thebibliography}
\end{document}